\documentclass[twocolumn]{aastex63}
\usepackage{longtable}
\usepackage{apjfonts}
\usepackage{mathrsfs}
\usepackage{booktabs}
\usepackage{amsmath}
\usepackage{natbib}
\usepackage{bm}

\newcommand{\xmm}{\hbox{\it XMM-Newton\/}}
\newcommand{\chandra}{{\it Chandra\/}}
\newcommand{\swift}{{\it Swift\/}}

\newcommand{\xray}{\hbox{\it \rm X-ray\/}}

\begin{document}
\title{On the Observational Difference Between the Accretion 
Disk-Corona Connections among Super- and \hbox{Sub-Eddington} Accreting Active Galactic Nuclei}

\author{Hezhen~Liu}
\affiliation{School of Astronomy and Space Science, Nanjing University, Nanjing, Jiangsu 210093, China; bluo@nju.edu.cn}
\affiliation{Department of Astronomy \& Astrophysics, 525 Davey Lab,
The Pennsylvania State University, University Park, PA 16802, USA}
\affiliation{Key Laboratory of Modern Astronomy and Astrophysics (Nanjing University), Ministry of Education, Nanjing 210093, China}

\author{B. Luo}
\affiliation{School of Astronomy and Space Science, Nanjing University, Nanjing, Jiangsu 210093, China; bluo@nju.edu.cn}
\affiliation{Key Laboratory of Modern Astronomy and Astrophysics (Nanjing University), Ministry of Education, Nanjing 210093, China}

\author{W. N. Brandt}
\affiliation{Department of Astronomy \& Astrophysics, 525 Davey Lab,
The Pennsylvania State University, University Park, PA 16802, USA}
\affiliation{Institute for Gravitation and the Cosmos,
The Pennsylvania State University, University Park, PA 16802, USA}
\affiliation{Department of Physics, 104 Davey Lab, The Pennsylvania State University, University Park, PA 16802, USA}

\author{Michael S. Brotherton}
\affiliation{Department of Physics and Astronomy, University of Wyoming, Laramie, WY 82071, USA}

\author{S.~C.~Gallagher}
\affiliation{Department of Physics \& Astronomy and Institute for Earth and Space Exploration, The University of Western
  Ontario, London, ON, N6A 3K7, Canada}

\author{Q. Ni}
\affiliation{Department of Astronomy \& Astrophysics, 525 Davey Lab,
The Pennsylvania State University, University Park, PA 16802, USA}
\affiliation{Institute for Gravitation and the Cosmos, The Pennsylvania State University, University Park, PA 16802, USA}

\author{Ohad~Shemmer}
\affiliation{Department of Physics, University of North Texas, Denton, TX 76203, USA}

\author{J. D. Timlin~III}
\affiliation{Department of Astronomy \& Astrophysics, 525 Davey Lab,
The Pennsylvania State University, University Park, PA 16802, USA}
\affiliation{Institute for Gravitation and the Cosmos,
The Pennsylvania State University, University Park, PA 16802, USA}

\begin{abstract}
We present a systematic \xray\ and multiwavelength study of a 
sample of 47 active galactic nuclei (AGNs) with reverberation-mapping 
measurements. This sample includes 21 \hbox{super-Eddington} 
 accreting AGNs and 26 \hbox{sub-Eddington} accreting AGNs.
 Using high-state observations with simultaneous \xray\ and 
 UV/optical measurements, we investigate whether 
 \hbox{super-Eddington} accreting AGNs exhibit different 
 accretion \hbox{disk-corona} connections compared to 
 \hbox{sub-Eddington} accreting AGNs. We find tight correlations 
 between the \hbox{X-ray-to-UV/optical} spectral slope parameter 
 ($\alpha_{\rm OX}$) and the monochromatic luminosity at 
 $2500~{\textup{\AA}}$ ($L_{\rm 2500~{\textup{\AA}}}$) for both the 
 super- and \hbox{sub-Eddington} subsamples. The best-fit 
 $\alpha_{\rm OX}\textrm{--}L_{\rm 2500~{\textup{\AA}}}$ relations 
 are consistent overall, indicating that super-Eddington accreting 
 AGNs are not particularly \xray\ weak in general 
 compared to sub-Eddington
 accreting AGNs. We find dependences of $\alpha_{\rm OX}$ on both the 
 Eddington ratio ($L_{\rm Bol}/L_{\rm Edd}$) and black hole mass
 ($M_{\rm BH}$) parameters for our full sample. A multi-variate linear 
 regression analysis yields $\alpha_{\rm OX}=-0.13 {\rm log}(L_{\rm 
 Bol}/L_{\rm Edd})-0.10 {\rm log}M_{\rm BH}-0.69$, with a scatter 
 similar to that of the $\alpha_{\rm OX}\textrm{--}L_{\rm
 2500~{\textup{\AA}}}$ relation.
 The hard (rest-frame $>2\rm ~keV$) \xray\ photon index ($\Gamma$) is
 strongly correlated with $L_{\rm Bol}/L_{\rm Edd}$ for the full 
 sample and the \hbox{super-Eddington} subsample, but these two
 parameters are not significantly correlated for the \hbox{sub-Eddington} subsample.
 A fraction of \hbox{super-Eddington} accreting AGNs show 
strong \xray\ variability, probably due to small-scale gas
absorption, and we highlight the importance of employing 
high-state (intrinsic) \xray\ radiation to study the accretion 
disk-corona connections in AGNs.

\end{abstract}

\section{Introduction}\label{sec:intro}
Active galactic nuclei (AGNs) are powered by 
accretion onto central massive black holes (BHs). 
The inflowing gas naturally forms an accretion disk, producing 
luminous emission mainly in the optical and UV bands.
Evidence indicates that the primary AGN X-ray emission
is produced via Compton up-scattering of the \hbox{accretion-disk}
UV/optical photons by energetic electrons in a compact corona 
surrounding the BH \citep[e.g.,][]{Sunyaev1980,Haardt1993}. 
  The emitted \xray\ photon spectrum is well
   described with a power-law continuum with a form of 
   $N(E) \propto E^{-\Gamma}$. 
   The photon index ($\Gamma$) of the spectrum 
   is expected to depend on coronal parameters such as 
   the electron temperature and optical depth
   \citep[e.g.,][]{Rybicki1986,Haardt1991,Haardt1993}.

As would be expected from the paradigm described above, 
observations have revealed evidence 
 that the accretion disk and \xray\ corona are connected.
 First, there is a strong correlation between AGN UV/optical 
  and \xray\ luminosities, which is often parameterized as a relation 
  between the \hbox{optical-to-\xray} \hbox{power-law} slope parameter 
  $\alpha_{\rm OX}$\footnote{$\alpha_{\rm OX}$ is defined as 
$\alpha_{\rm OX}=0.3838{\rm log}(L_{\rm 2~keV}/L_{\rm
 2500~{\textup{\AA}}})$, where ${\sc L_{\rm 2~keV}}$ and $L_{\rm 
 2500~{\textup{\AA}}}$ are the monochromatic luminosities at 
 rest-frame 2~keV and $2500~{\textup{\AA}}$, respectively.} 
and $L_{\rm 2500~{\textup{\AA}}}$ across a broad range in UV 
luminosity \citep[e.g.,][]
 {Avni1982,Avni1986,Kriss1985,Wilkes1994,Vignali2003,Strateva2005,
Steffen2006,Just2007,Gibson2008a,Green2009,Lusso2010,Grupe2010,
Jin2012,Laha2018,Chiaraluce2018}.
This correlation can also be described as the dependence of 
${\sc L_{\rm 2~keV}}$ on $L_{2500~{\textup{\AA}}}$ \citep[e.g.,][]
{Tananbaum1979,Avni1982,Avni1986,Wilkes1994,Strateva2005,Steffen2006,
Just2007,Lusso2010,Lusso2016,Chiaraluce2018}. 
These correlations suggest that some mechanisms 
\citep[e.g.,][]{Merloni2003,Liu2002,Jiang2019,Arcodia2019,Cheng2020} 
must be in place to regulate the energetic interactions between the 
accretion disk and the corona. A well established $\alpha_{\rm OX}
\textrm{--} L_{2500~{\textup{\AA}}}$ or ${\sc L_{\rm 2~keV}} 
\textrm{--} L_{2500~{\textup{\AA}}}$ relation  
 can be used to identify AGNs emitting unusually weak or strong \xray\ 
 emission, relative to their UV luminosities \citep[e.g.,][]
 {Gibson2008a,Miller2011}. A well-calibrated ${\sc L_{\rm 2~keV}} 
\textrm{--} L_{2500~{\textup{\AA}}}$ relation 
can also be used to estimate cosmological parameters \citep[e.g.,][]
 {Lusso2016,Lusso2017,Bisogni2017}.
 
 Another remarkable piece of evidence showing the \hbox{disk-corona} 
 connection is the significant positive 
 correlation between the hard \xray\ photon index and Eddington ratio 
 ($L_{\rm Bol}/L_{\rm Edd}$, where $L_{\rm Bol}$ is the 
 bolometric luminosity and $L_{\rm Edd}$ 
is the Eddington luminosity; e.g.,
 \citealt{Shemmer2006,Shemmer2008,Jin2012,Risaliti2009,Brightman2013,
Trakhtenbrot2017}).
 A possible interpretation of this relationship is that
 in a higher $L_{\rm Bol}/L_{\rm Edd}$ system the enhanced UV/optical 
 emission from the accretion disk results in 
more effective Compton cooling of the corona, decreasing its 
temperature and/or optical depth, which leads to the softening of the \xray\ spectrum (a larger value of $\Gamma$; e.g.,
 \citealt{Haardt1991,Haardt1993,Pounds1995,Fabian2015,Cheng2020}).

The AGN accretion-disk properties likely
depend on the accretion rate (usually represented by $L_{\rm Bol}/L_{\rm Edd}$).
Typical AGNs with $0.01\la L_{\rm Bol}/L_{\rm Edd} \la 0.3$ are 
expected to have optically thick geometrically thin accretion disks 
 \citep[e.g.,][]{Shakura1973}.
 At higher accretion rates ($L_{\rm Bol}/L_{\rm Edd}\ga 0.3$), 
the disk likely becomes geometrically thick, as 
shown in the slim disk model \citep[e.g.,][]
{Abramowicz1980,Abramowicz1988,Wang2003,Wang2014b} and numerical 
simulations \citep[e.g.,][]
{Ohsuga2011,Jiang2014,Jiang2016,Jiang2017,Kitaki2018,Skadowski2014}.
The radiative efficiency of the thick disk is probably reduced 
compared to the thin disk, mainly due to the 
\hbox{photon-trapping} effect, although the values of the radiative 
efficiency predicted by numerical simulations are much larger than 
that of the analytic slim disk model, close to that of the thin disk.
The spectral energy distribution (SED) of the thick disk is 
 expected to be different from that of the thin disk, and
 the thick disk may produce enhanced extreme UV 
 (EUV; $\sim 100\textrm{--}1200~\textup{\AA}$) radiation 
  \citep[e.g.,][]{Castell2016,Kubota2018}.
However, this difference is hard to resolve observationally due to 
the lack of EUV data.
 As a result of the changing disk structure, both the nature of 
 the coronae and the accretion disk-corona connections may be 
 different between super- and sub-Eddington accreting AGNs.
For example, the \hbox{photon-trapping} effect may influence the 
energy transport from the accretion disk to the corona, changing 
the Compton cooling efficiency of hot electrons in the corona.

If \hbox{super-Eddington} accreting AGNs have  
different coronae and/or accretion disk-corona connections 
compared to \hbox{sub-Eddington} accreting AGNs, this probably 
manifests in altered $\alpha_{\rm OX}(L_{\rm 2~keV})\textrm{--}
L_{\rm 2500~{\textup{\AA}}}$
 and $\Gamma\textrm{--}L_{\rm Bol}/L_{\rm Edd}$ relations.
 Previous studies of these relations usually targeted samples 
 including both super- and \hbox{sub-Eddington} accreting
 AGNs, and the two groups were not separated. 
There have been some suggestions that \hbox{super-Eddington} 
  accreting AGNs may show relatively weak \xray\ emission,
  deviating from the $\alpha_{\rm OX}(L_{\rm 2~keV})\textrm{--}L_{\rm 
2500~{\textup{\AA}}}$ relation for \hbox{sub-Eddington} accreting 
AGNs. For example, \xray\ investigations of \hbox{intermediate-mass}
 BH (IMBH) candidates with high Eddington ratios suggested  
 that a number of IMBHs appear to have suppressed \xray\ emission 
 \citep[e.g.,][]{Greene2007,Dong2012}. However, weak X-ray
 emission observed from \hbox{super-Eddington} accreting AGNs might
 not be intrinsic, and it may instead be caused by X-ray absorption. Some narrow-line Seyfert~1 (NLS1) galaxies and quasars with 
 high accretion rates have been found to show extreme (by factors of 
 $>10$) X-ray variability without coordinated   
 UV/optical variability, and they are significantly 
X-ray weak in the low X-ray states; their extreme X-ray weakness 
is probably related to the shielding of the corona
 by a thick inner accretion disk and its associated outflow
 \citep[e.g.,][and 
 references therein]{Liu2019,Ni2020}. In addition, although there have 
 been some studies of the $\Gamma\textrm{--}L_{\rm Bol}/L_{\rm Edd}$ 
 relation for AGNs with high Eddington ratios \citep[e.g.,][]
 {Ai2011,Kamizasa2012}, they have not compared systematically the
 difference between the relations among super- and 
 \hbox{sub-Eddington} accreting AGNs.

In this work, we utilize an AGN sample with reverberation 
mapping (RM) measurements to investigate systematically 
the observational difference between the accretion 
disk-corona connections among super- and \hbox{sub-Eddington} 
accreting AGNs. The RM technique is considered to provide the 
 most accurate measurements of BH masses for broad emission-line AGNs  
   \citep[e.g.,][]{Blandford1982,Peterson1993}. 
Successful RM studies have been carried out for nearly 200 AGNs 
   \citep[e.g.,][]{Peterson1998,Peterson2002,Peterson2004,Kaspi2000,Kaspi2007,
Bentz2008,Bentz2009,Denney2009,Barth2013,Barth2015,Rafter2011,Rafter2013,
Du2014,Du2015,Du2016,Du2018,Wang2014a,Shen2016,Jiang2016,
Fausnaugh2017,Grier2012,Grier2017,Grier2019}.
   Most of these campaigns do not target \hbox{super-Eddington} 
   accreting AGNs. A recent campaign targeting \hbox{super-Eddington} 
   accreting massive black holes (SEAMBHs) has identified about 24 
   candidates, which increases significantly the number of 
   \hbox{super-Eddington} 
   accreting AGNs with RM measurements 
   \citep{Du2014,Du2015,Du2016,Du2018,Wang2014a}.
   The majority of reported RM AGNs are nearby bright sources that 
   have sensitive \xray\ coverage from archival \xmm\ 
   \citep{Jansen2001}, \chandra\ \citep{Weisskopf1996}, and \swift\ 
   \citep{Gehrels2004} observations. For \xmm\ and \swift\ 
   observations, simultaneous UV/optical measurements by the same 
   satellites are available.

The paper is organized as follows. In Section~\ref{sec:data},
we describe the sample-selection procedure, the observational data,
and the methods of data reduction and analysis. 
In Section~\ref{sec:result},
we present the statistical analysis of the correlations between 
$\alpha_{\rm OX}$ ($L_{\rm 2~keV}$) and $L_{\rm 
2500~{\textup{\AA}}}$, $\Gamma$ and $L_{\rm Bol}/L_{\rm Edd}$,
for our super- and \hbox{sub-Eddington} accreting AGN subsamples.   
In Section~\ref{sec:discuss},
   we discuss the implications of the correlations, and we explore the
   dependence of $\alpha_{\rm OX}$ on 
   both $M_{\rm BH}$ and $L_{\rm Bol}/L_{\rm Edd}$.
   We summarize and present future prospects in 
   Section~\ref{sec:sum}. Throughout this paper,
we use J2000 coordinates and a cosmology with
$H_0=67.4$~km~s$^{-1}$~Mpc$^{-1}$, $\Omega_{\rm M}=0.315$,
and $\Omega_{\Lambda}=0.686$ \citep{Planck2018}.
All errors are quoted at a $68\%$ ($1\sigma$) confidence level.

\section{Sample Properties and Data Analysis} 
\label{sec:data}
\subsection{Sample Selection}
\label{subsec:sample}
Our RM AGN sample was selected from the RM AGNs compiled 
by the SEAMBH collaboration \citep{Du2015,Du2016,Du2018}, which
contains 25 targets (including 24 \hbox{super-Eddington} 
accreting AGN candidates) in the SEAMBH campaign and 50
 AGNs studied in previous RM work. 
We selected sample objects from these 75 AGNs based on the following 
considerations: 
 
(1) We selected radio-quiet AGNs, since radio-loud AGNs may have excess \xray\ emission linked to the radio jets and other processes \citep[e.g.,][]{Miller2011,Zhu2020}.
We searched for radio information from the literature 
\cite[e.g.,][]{Kellermann1989,Wadadekar2004,Sikora2007} and the NASA/IPAC Extragalactic Database (NED),\footnote{https://ned.ipac.caltech.edu.} and we discarded five radio-loud AGNs.

(2) Since we aim to explore the intrinsic \xray\ and UV/optical 
properties of our sample, objects affected by heavy absorption in the 
\xray\ and UV/optical bands were excluded. We discarded 11 
objects in total, nine of which are well-studied objects 
(PG~$1700+518$, 
Mrk~486, NGC~4051, NGC~4151, NGC~3516, NGC~3227, NGC~3783, 
Mrk~5273, and \hbox{MCG--6-30-15}) with \xray\ emission affected by ionized and/or neutral absorption related to outflows, and some 
of them also show continuum reddening or broad absorption 
lines (BALs) in their UV spectra \citep[e.g.,][]
{Pounds2004,Kraemer2005,Kraemer2006,Kraemer2012,Cappi2006,Ballo2008,Ballo2011,Turner2011,
Beuchert2015,Pahari2017,Dunn2018}.
A \hbox{super-Eddington} accreting AGN, Mrk~202,  
shows an unusually flat 2--10~keV spectrum and an absorption 
feature prominent in the $<2$~keV spectrum, so that it was also 
excluded. In addition, a "changing-look" AGN, Mrk~590, has changed
its optical spectral type from Type 1 to Type \hbox{1.9--2}, 
with dramatic variation in the UV/optical and \xray\ fluxes
\citep[e.g.,][]{Denney2014}. It is expected to emit intrinsic 
\xray\ and UV/optical radiation in the high state when it 
was classified 
as a Type 1 Seyfert galaxy. However, there is no \xray\ 
observation available during the high state. We thus excluded this 
AGN from our study.  

There are three additional AGNs that have at one time shown 
absorption features in the \xray\ and/or UV/optical bands.
A transient absorption event has been detected in Mrk~766,
but its \hbox{high-state} \xray\ spectrum does not show any
absorption features \citep[e.g.,][]{Risaliti2011,Liebmann2014}. 
We thus retained this object 
and used its high-state observational data in our analysis.
The 2013 \xmm\ observation of NGC~5548 revealed that it was
obscured by clumpy gas not seen before, which blocks a large 
fraction of its soft \xray\ emission and causes UV BALs \citep[e.g.,][]{Kaastra2014,Cappi2016}.
We checked its \hbox{high-state} \xray\ and UV/optical data obtained 
from the 2002 \xmm\ observation, and we found that the simultaneous 
\xray\ and UV/optical fluxes are free from absorption.  
Its 2002 UV spectrum also shows no absorption features 
\citep[e.g.,][]{Kaastra2014}. We thus also retained this object in 
our study and analyzed its 2002 \xmm\ observation. 
 Moreover, a \hbox{super-Eddington} accreting AGN, 
 IRAS F$12397+3333$, was affected by ionized absorption with flux
 deficiencies prominent around \hbox{0.7--1~keV}, while its
 2--10~keV spectrum is hardly 
affected by the ionized absorption \citep{Dou2016}. Its steep Balmer 
decrement indicates intrinsic reddening in the 
UV/optical \citep[e.g.,][]
{Du2014}. Considering that its hard \xray\ spectrum is not 
affected by absorption, we still included this object in our study 
and applied a reddening correction to its UV/optical data (see Section~\ref{subsec:spec} and Appendix for details).
 
(3) The selected AGNs have good archival \xmm, \chandra, or 
 \swift\ coverage. Since we focused on the hard 
 (\hbox{rest-frame} $>2$~keV)
 \hbox{X-rays} (see Section~\ref{subsec:spec} below), 
   these objects are required to be
   significantly detected in the \hbox{rest-frame} $>2$~keV band 
   with signal-to-noise (S/N) ratios larger than 6. 
The \xmm\ data have the highest priority since they have 
high-quality \xray\ spectral data and simultaneous 
UV/optical data. 
For AGNs without \xmm\ data, we used   
\chandra\ observations if available, otherwise, \swift\ 
observations are used.
There are six AGNs without archival \xray\ data, of which 
four are Sloan Digital Sky Survey Release 7 (SDSS-DR7) quasars in the
 SEAMBH campaign. 
In addition, there are six SDSS quasars in the SEAMBH campaign    
serendipitously detected by \swift\ with low S/N ratios.
We thus discarded these 12 AGNs. 

Our final sample consists of 47 RM AGNs, which are referred to as the 
full sample. 
From these objects we identified 21 \hbox{super-Eddington} accreting 
AGNs that were classified on the basis of normalized accretion rate of 
$\dot{\rm \mathscr{M}}\ge 3$, following the approach of the 
SEAMBH campaign \citep{Wang2014a,Du2015}. 
The normalized accretion rate is defined as 
$\dot{\rm \mathscr{M}}=\dot{M}
c^2/L_{\rm Edd}$, where $\dot{M}$ is the mass accretion rate. 
We measured $\dot{\rm \mathscr{M}}$ based on the \cite{Shakura1973} 
standard thin disk model (see Section~\ref{subsec:mdot}).
For super-Eddington accreting AGNs,  
$\dot{\rm \mathscr{M}}$ may be a better indicator of the accretion 
rate, rather than $L_{\rm Bol}/L_{\rm Edd}$, because 
their bolometric luminosities may be saturated due to the  
photon-trapping effect \citep[e.g.,][]{Wang2014a}. 
The criterion of $\dot{\rm \mathscr{M}}\ge 3$ 
 corresponds to $L_{\rm Bol}/L_{\rm Edd}\ge 0.3$ for a typical 
 radiative efficiency of $\eta=0.1$.
 We refer to the 21 super-Eddington accreting AGNs 
 as the \hbox{super-Eddington} 
 subsample. The remaining 26 objects constitute the 
  \hbox{sub-Eddington} subsample. 
  
\begin{figure}
 \centering
  \includegraphics[scale=0.44]{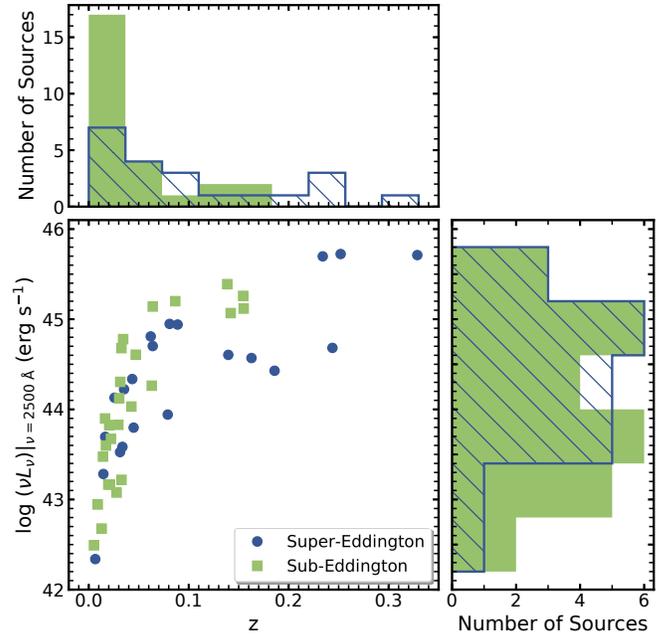}
\caption{Redshift vs. $2500~{\textup{\AA}}$ luminosity for the \hbox{super-Eddington} subsample (blue circles) and the \hbox{sub-Eddington} 
subsample (green squares). The right panel shows the distribution 
of $2500~{\textup{\AA}}$ luminosity, and the upper panel shows the 
distribution of redshift. The \hbox{super-Eddington} (\hbox{sub-Eddington}) 
subsample is represented by the hatched blue (filled green)
histograms.
} 
\label{fig-dis}
\end{figure}

  Table~\ref{tbl1} lists the basic 
  properties of our sample AGNs including $5100~\textup{\AA}$ luminosities, H$\beta$ FWHMs, and BH masses ($M_{\rm BH}$) 
  and associated measurement uncertainties, which were adopted from
    \cite{Du2015,Du2016,Du2019}. 
    We note that there may be considerable systematic uncertainties on 
    the RM BH masses (see Section~\ref{subsec:dis-gamma} below for 
     discussion), which are 
    difficult to quantify. We thus only take into account the 
    measurement uncertainties in the following 
    analyses. For the full sample, the measurement uncertainties 
    on $M_{\rm BH}$ range from 0.03 to 0.40 dex, with a median value of 0.13 dex. 
    Figure~\ref{fig-dis} shows the distribution of our sample 
AGNs in the redshift versus $2500~{\textup{\AA}}$ luminosity plane. For all sample objects 
(except SDSS~J$085946+274534$), the $2500~{\textup{\AA}}$ luminosities
 were derived from the UV/optical 
 data that were observed simultaneously with the \xray\ data 
 (see detailed analyses in the following subsections). 
The distributions of 
the $2500~{\textup{\AA}}$ luminosities for the super-
and \hbox{sub-Eddington} subsamples are comparable.

 Some of our sample objects are well-studied AGNs with multiple 
 \xray\ observations. For these objects, we used their
high-state observational data, which likely correspond to 
the intrinsic \xray\ emission. 
 There are three \hbox{super-Eddington} subsample objects (Mrk~335, 
PG~$1211+143$, and PG~$0844+349$) displaying extreme
 \xray\ variability by factors of larger than 10 \citep[e.g.,][]
{Gallagher2001,Grupe2007a,Bachev2009,Gallo2011,Grupe2012,Gallo2018}.
Their high-state \xray\ observations reveal 
\xray\ fluxes consistent with the levels expected from their 
UV/optical emission, likely reflecting their intrinsic 
\xray\ properties (see notes on individual sources in Appendix). 
Moreover, we included two "changing-look" AGNs with 
\hbox{sub-Eddington} 
accretion rates (NGC~2617: \citealt{Shappee2014,Giustini2017}; Mrk~1310: Luo B. et al. in preparation; 
see details in Appendix). They have undergone changes in their 
optical spectral types and UV/optical and \xray\ fluxes.  
We used their observational data in the historical high states, when 
they exhibit strong broad optical emission lines and the brightest 
\xray\ and UV/optical luminosities. 
A list of the \xray\ observations used in 
this study is presented in Table~2.

\subsection{XMM-Newton Observations}
\label{subsec:xmm}
{\it XMM-Newton} data are available for 37 sample AGNs. 
Except for one AGN, SBS 1116+583A, all these AGNs were targets of the 
corresponding \xmm\ observations. 
Simultaneous \xray\ and UV/optical data were obtained 
from the European Photon Imaging Camera (EPIC) PN
 \citep{Struder2001} and MOS \citep{Turner2001} detectors,
 and the Optical Monitor \citep[OM;][]{Mason2001}.
 We processed the data using the \xmm\ Science Analysis 
 System (SAS v.16.0.0), and the latest calibration files were 
 applied. For the \xray\ analysis, we only used the EPIC PN data.
 The task {\it epproc} was first used to reduce the PN data and 
 create the calibrated event lists. 
  Bad or hot pixels were removed from the event lists, and 
high-background flares were checked and filtered according to the 
standard criteria. Only single and double events (PATTERN$\le 4$ and 
FLAG=0) were selected.
A circular region with a default radius of 35\arcsec\ was 
used to extract the source spectrum.  
Each data set was inspected for pile-up by running the task 
 {\it epatplot}. For nine sources with detected 
 pile-up (see Table~\ref{tbl2}), 
 an inner circular region with a radius of 
 5--10\arcsec\ was discarded from the source extraction region. 
These are bright X-ray sources, and the extracted spectra still have high 
 quality, allowing robust spectral fitting.
 The background spectrum was extracted from 
 a nearby source-free circular region with a radius of 
 50--100\arcsec\ on the same CCD chip. 
The response matrix and ancillary response function files 
were created using the tasks {\it rmfgen} and {\it arfgen}, 
which account for the point spread function correction of the  
source-extraction region.
The final PN spectrum was grouped with a minimum number of one count 
per bin using the task {\it specgroup}.

We analyzed OM data mainly for deriving flux 
densities at the rest-frame $2500~{\textup{\AA}}$ band.
OM imaging data are available for all sample objects, 
except for Mrk~335 with only UV grism data in its selected 
observation. There are six OM filters, including three optical filters 
(V, B, U with central wavelengths of 5430~${\textup{\AA}}$, 
4500~${\textup{\AA}}$, 3440~${\textup{\AA}}$) and three UV filters
 (UVW1, UVM2, UVW2 with central wavelengths of 2910~${\textup{\AA}}$, 
 2320~${\textup{\AA}}$, 2110~${\textup{\AA}}$).  
We processed the OM filter data using the pipeline 
{\it omchian}. Point sources and extended sources  
were automatically identified as part of the pipeline.
Source fluxes and magnitudes were extracted from the SWSRLI 
files, and we adopted the mean magnitudes and fluxes of all the 
exposure segments for each filter.
For Mrk 335, the OM UV grism data were processed with 
the pipeline {\it omghian}. The pipeline generated 24 calibrated 
spectra, and the spectra cover a wavelength range 
of $1800\textrm{--}3600~{\textup{\AA}}$.
The fluxes of the OM filters and the mean grism spectra were then 
corrected for Galactic extinction using the extinction law of \cite{Cardelli1989}.
Table~\ref{tbl1} lists the mean values of Galactic extinction $E_{\rm 
B-V}$ \citep{Schlegel1998} that were obtained from the NASA/IPAC 
Infrared (IR) Science Archive. \footnote{https://irsa.ipac.caltech.edu/applications/DUST/.} 

We utilized the OM UV-filter flux densities 
to derive the $2500~{\textup{\AA}}$ flux densities. 
Our sample objects are bright, and at least in the UV-filter 
images, they were identified as point sources. Therefore, host-galaxy 
contamination in the \hbox{UV-filter} fluxes should be mild 
(see \citealt{{Grupe2010}} for discussion regarding the \swift\ 
UV/optical photometric data), and we did not correct the source fluxes 
and magnitudes for any host-galaxy contamination. 
Eleven out of the 36 objects with OM imaging data were 
observed with three UV filters, and another 14 objects were observed 
with two filters. We derived their $2500~{\textup{\AA}}$ flux 
densities by fitting a \hbox{power-law} model to the observed data. 
For eight objects observed with only one UV filter, 
the $2500~{\textup{\AA}}$ 
flux densities were extrapolated assuming a power-law slope of 
$\alpha_{\nu}=-0.44$ \citep[e.g.,][]{Vanden2001}.  
 For three other objects without \hbox{UV-filter} data, 
 NGC~5548, NGC~6814 and PG~$0844+349$, their $2500~{\textup{\AA}}$
  flux densities were extrapolated 
 from the \hbox{U-filter} flux densities assuming the same power-law 
 slope of $\alpha_{\nu}=-0.44$.  
 Finally, the $2500~{\textup{\AA}}$ flux density of Mrk~335 was 
 measured from the mean grism spectrum. The UV-filter information  
and the derived $L_{2500~\textup{\AA}}$ values are listed in Table~\ref{tbl2}.

\subsection{Chandra Observations}
\label{subsec:chandra}
We used \chandra\ data for only one object, SDSS~J$085946+274534$. 
It was observed as a target on 2004 December 25.
The observational data were analyzed using 
the \chandra\ Interactive Analysis of Observations (CIAO; v4.11) 
tools. A new level 2 event file was generated using 
 the {\sc chandra\_repro} script, and 
  \hbox{high-background} flares were filtered by running the {\sc 
 deflare} script with an iterative 3$\sigma$ clipping algorithm.
 A 0.5--7~keV image was then constructed by running the 
 {\sc dmcopy} script. 
 The {\sc specextract} tool was used to extract and group 
 spectra (with at least one count per bin), 
 and to generate the response matrix and ancillary 
 response function files.
The \hbox{source-extraction} region is  
a circular region with a radius of 3\arcsec, centered on the \xray\ source position
  detected by the automated \hbox{source-detection} tool 
  {\sc wavdetect}.
  An annulus region centered on the \xray\ 
  source position with a 10\arcsec\ inner radius and a 30\arcsec\ 
  outer radius was chosen as the background-extraction region. 
 The extracted source spectrum was grouped with a minimum number 
 of one count per bin.

 There are no simultaneous UV/optical data available for this 
 \chandra\ object. We interpolated 
 its Near-UV (NUV) flux density, observed on 2006 February 18, from 
 {\it Galaxy~Evolution~Explorer}
 \cite[{\it GALEX};][]{Martin2005} and SDSS {\it u}-band flux density, 
 observed on 2004 April 17, to derive the $2500~{\textup{\AA}}$ flux 
 density.

\subsection{Swift Observations}
\label{subsec:swift}
We used \swift\ data for nine AGNs. Simultaneous \xray\ and 
UV/optical data are available from  
the \xray\ Telescope (XRT; \citealt{Burrows2005}) and the UV-Optical
Telescope (UVOT; \citealt{Roming2005}). For all observations, 
the XRT was operated in the Photon Counting (PC) mode 
\citep{Hill2004}. 
The data were reduced with the task {\it xrtpipeline} version 0.13.4, 
which is included in the HEASOFT package 6.25. The XRT data were 
not affected by photon pipe-up given the low count rates 
($0.02\textrm{--}0.14\ \rm counts\ s^{-1}$) of these nine AGNs.
For each source, the source photons were extracted using the task 
{\it xselect} version 2.4, from a circular 
region with a radius of 47\arcsec. The background spectrum was
 extracted from a nearby source-free circular region with a radius of 
 100\arcsec. The ancillary response function file was generated by 
 {\it xrtmkarf}, and the standard photon redistribution matrix file 
 was obtained from the CALDB. We grouped the spectra using 
 {\it grppha} such that each bin contains at least one photon.

The UVOT has a similar set of filters (V, B, U, UVW1, UVM2, UVW2 
with central wavelengths of 5468~${\textup{\AA}}$, 4392~${\textup{\AA}}$, 
3465~${\textup{\AA}}$, 2600~${\textup{\AA}}$, 2246~${\textup{\AA}}$, 
and 1928~${\textup{\AA}}$) to the OM. Similarly, we used mainly
 the UV-filter data to derive the $2500~{\textup{\AA}}$ flux 
 densities. Among these nine \swift\ AGNs, six  
 were observed with one UV filter, one was observed with two 
 UV filters, and two were observed with three UV filters. 
 The data from each segment in each filter were co-added using the 
task {\it uvotimsum} after aspect correction.
Source counts were selected from a circular region with a radius of 
5\arcsec, centered on the source position determined by the task 
{\it uvotdetect} \cite{Freeman2002}. A nearby source-free region with
 a radius of 20\arcsec\ was used to extract background counts.
Source magnitudes and fluxes in each UVOT filter were then 
computed using the task {\it uvotsource}, and these 
data were corrected for Galactic extinction.
We derived the $2500~{\textup{\AA}}$ flux densities 
 following the same procedure as used for the OM photometric data. 
 
\subsection{X-ray Spectral Analysis}
\label{subsec:spec} 
X-ray spectral analysis was performed with XSPEC (v12.10.1; 
\citealt{Arnaud1996}).
All spectra were grouped with at least one count per bin, 
and the Cash statistic (CSTAT; \citealt{Cash1979}) 
was applied in parameter estimation; the W statistic was actually 
used because the background 
spectrum was included in the spectral fitting.\footnote{
https://heasarc.gsfc.nasa.gov/xanadu/xspec/imanualXSappendixStatistics.html.} 
Since the aim of this study is to obtain 
properties of the intrinsic coronal \xray\ radiation, we focused 
only on the rest-frame $> 2~\rm keV$ energy band,
where the observed \xray\ emission is less affected by contamination 
from a potential soft excess component and intrinsic absorption 
\citep[e.g.,][]{Shemmer2006,Shemmer2008,Risaliti2009,Brightman2013}.
The \hbox{rest-frame} $< 2~\rm keV$ spectral data were also analyzed
  with basic phenomenological models, in order to construct \hbox{broad-band} 
  SEDs and provide more reliable estimates of bolometric luminosities for our 
 sample objects (see Section~\ref{subsec:bol} below).

\begin{figure}
\centerline{
\includegraphics[scale=0.58]{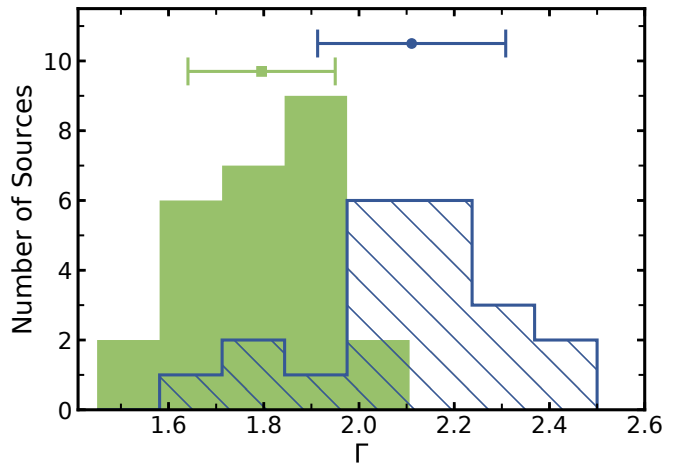}
}
\caption{Distribution of the \xray\ spectral photon indices for 
the \hbox{super-Eddington} (hatched blue histograms) and \hbox{sub-Eddington} 
(filled green histograms) subsamples. The two points with error 
bars on the top show the mean values and $1\sigma$ deviations 
for the two subsamples. The two subsamples are 
clearly distinct, with \hbox{super-Eddington} accreting AGNs 
showing steeper (softer) hard \xray\ spectra.}
\label{fig_gamma}
\end{figure}

We adopted a power-law model modified by Galactic
 absorption ({\sc phabs*zpowerlw}) to 
fit the rest-frame \hbox{$> 2~\rm keV$} spectra, corresponding to the
\hbox{observed-frame} $2/(1+z)\textrm{--} 10~\rm keV$ spectra 
for \xmm\ and \swift\ observations and the observed-frame 
$2/(1+z)\textrm{--} 7~\rm keV$ spectra for the \chandra\ 
observation. We visually inspected whether there are strong iron K 
lines in the spectra. For 25 AGNs in which the iron K lines were 
detected, the rest-frame $5.5\textrm{--}7.5~\rm keV$ spectra were 
discarded in the spectral fitting.
The Galactic neutral hydrogen column density toward each source 
was fixed at the value from \cite{Dickey1990}.
For all objects, the statistics of the best-fit models are acceptable
(see Table~\ref{tbl2}), and the fitting residuals are distributed 
close to zero without any apparent systematic excesses/deficiencies.
As shown in Figure~\ref{fig_gamma}, the photon indices ($\Gamma$) of 
the full sample span a range of $1.50\textrm{--}2.46$ with a median 
value of 1.94.
In general, the \hbox{super-Eddington} subsample has larger (softer) 
photon indices than the \hbox{sub-Eddington} subsample.

Since we have discarded \xray\ and UV/optical absorbed AGNs  
from our sample, and the high-state data for objects 
with multiple observations were used, it is likely that intrinsic 
 absorption has little impact on the derived \xray\ properties of 
 our sample. To confirm this, we checked for the presence of 
intrinsic absorption in each object by adding an intrinsic neutral
 absorption component 
({\sc zphabs} model in XSPEC) to the fitting model.  
For each object, the \hbox{best-fit} statistic did not 
change significantly, and the resulting column density is consistent 
with zero; an F test also indicated that the absorption 
component is likely not required.  
Moreover, we extrapolated the \hbox{data-to-model} ratios of the 
best-fit simple power-law model (not including the intrinsic 
absorption component) to the entire spectral energy range 
(\hbox{0.3--10}~keV for \xmm\ and \swift\ observations and 
\hbox{0.5--7} keV for the \chandra\ observation) to inspect 
whether absorption is present at rest-frame $<2~\rm keV$ energies. 
Only IRAS F$12397+3333$ shows flux deficiencies in the 
\hbox{$\approx$ 0.7--1~keV} band. It was reported by \cite{Dou2016} 
that the \hbox{0.3--10~keV} spectral fitting with a 
model including ionized absorption and soft excess components yields  
an intrinsic photon index of 2.2, which is consistent with the 
 photon index obtained by fitting the \hbox{2--10 keV} spectrum with 
 a simple power-law model. 
 Our simple \hbox{power-law} fitting of the \hbox{rest-frame} 
$>2~\rm keV$ spectrum also resulted in a consistent photon 
index of 2.14. We thus conclude that the rest-frame 
$>2~\rm keV$ spectrum of IRAS F$12397+3333$ is intrinsic. 
 Therefore, for all objects, we adopted the results from the spectral 
fitting performed with the simple power-law model modified with 
Galactic absorption. The flux densities at rest-frame
2~keV ($f_{\rm 2~keV}$) were then computed based on the 
best-fit models. The \hbox{best-fit} model parameters and fitting 
statistics are presented in Table~2.

\subsection{Estimation of Normalized Accretion Rates} 
\label{subsec:mdot}
The \cite{Shakura1973} standard thin disk model predicts a 
power-law SED in the form of $F_{\nu} \propto \nu^{1/3}$ from 
optical to NUV, and the monochromatic luminosity at a given wavelength
depends on the BH mass and accretion rate 
(e.g., Equation~5 of \citealt{Davis2011}).
Given the observed SED and the BH mass, the accretion
rate ($\dot{M}$ or $\dot{\rm \mathscr{M}}$) can be computed 
\citep[e.g.,][]{Davis2011,Netzer2013,Wang2014a}. 
We used the monochromatic luminosity at $2500~\textup{\AA}$ 
to calculate 
$\dot{\rm \mathscr{M}}$ following the expression:
\begin{equation}
\dot{\rm \mathscr{M}}= 4.82 (\ell_{44}/{\cos i})^{3/2} m_7^{-2},
\label{eq0}
\end{equation}
 where $\ell_{44} = 
 \nu L_{2500~\textup{\AA}}/ 10^{44}~\rm erg~s^{-1}$ is the 
  $2500~\textup{\AA}$ luminosity in units
 of $10^{44}~\rm erg~s^{-1}$, $m_7=M_{\rm BH}
 /10^7M_\odot$, and {\it i} is the inclination angle of the disk.
For Type 1 AGNs, the dependence of $\dot{\rm \mathscr{M}}$ on cos {\it i} is weak,\footnote{According to the unified model, Type 1 AGNs are 
 observed at relatively small inclination angles ($i$). For a typical $i$ range of $ 0\textrm{--}
 60\arcdeg$, cos {\it i} only varies by a factor of two ($0.5\textrm{--}1$). Thus, we adopted 
cos {\it i}$=0.75$ (see also discussions in \citealt{Du2014,Du2016,Wang2014a}).} and thus we 
adopted a median cos {\it i} value of 0.75. We adopted the 
 $2500~\textup{\AA}$ luminosity instead of the $5100~\textup{\AA}$ luminosity that is often used
 to calculate $\dot{\rm \mathscr{M}}$ in previous studies, because 
  the emission around $2500~\textup{\AA}$ is less 
affected by \hbox{host-galaxy} contamination that is  
significant for nearby moderate-luminosity AGNs.
 Moreover, for most sample objects, the
 $2500~\textup{\AA}$ luminosities were derived 
  from the UV/optical data that were observed simultaneously with 
  the \xray\ data,
  and thus the accretion-disk corona connections explored below are 
  free from any variability effects. For the full sample, the 
  $\dot{\rm \mathscr{M}}$ values span a range of $0.012$ to $530$, with a median value of $0.83$.
  The uncertainties on $\dot{\rm \mathscr{M}}$ are propagated 
  from the measurement uncertainties on $M_{\rm BH}$ and the 
  $2500~\textup{\AA}$ luminosities. 
  Table~\ref{tbl1} lists the 
  ${\rm log}\dot{\rm \mathscr{M}}$ values and their uncertainties.

  We note that Equation~\ref{eq0} likely also holds for 
  \hbox{super-Eddington} accreting 
  AGNs \citep[e.g.,][]{Du2016,Huang2020}, where the accretion disks 
  are expected to be geometrically thick. 
  Based on the self-similar solution of the slim 
  disk model \citep{Wang1999a,Wang1999b}, the radius of 
  the disk region emitting $2500~\textup{\AA}$ photons is 
  larger than the \hbox{photon-trapping} radius for our sample 
  objects.\footnote{Using the self-similar solution of the slim 
  disk model \citep{Wang1999a,Wang1999b} and Wien's law, we estimated 
  the radius of the disk region emitting $2500~\textup{\AA}$ photons 
to be $R_{2500}/R_{\rm g} \approx 2.1\times 10^3~m_7^{-1/2}$, and 
  the \hbox{photon-trapping} radius is given by 
  $R_{\rm trap}/R_{\rm g}\approx 450\ (\dot{\rm \mathscr{M}}/250)$,   
 where $R_{\rm g}=GM_{\rm BH}/c^2$ (see \citealt{Du2016,Cackett2020}). 
  Equation~\ref{eq0} holds provided that $R_{2500}>R_{\rm trap}$ 
  (i.e., $\dot{\rm \mathscr{M}}\la 1.2\times 10^3~m_7^{-1/2}$), and 
  this condition is met for all the objects in our sample.} 
  Observationally, studies of the SEDs of \hbox{super-Eddington} 
   accreting AGNs revealed that their UV/optical SEDs are well-fitted
   by the thin disk model, and the characteristics of the thick-disk
    emission likely emerge in the EUV (e.g.,
    \citealt{Castell2016,Kubota2018}). It is also supported by our 
    finding that the \hbox{high-luminosity} \hbox{super-Eddington} 
    accreting AGNs in our sample show UV/optical SEDs consistent with 
    those of typical quasars (see Section~\ref{subsec:sed}). 
    Furthermore, a recent accretion-disk reverberation mapping on a 
   super-Eddington accreting AGN, 
   Mrk~142, found that the UV/optical (rest-frame $\approx 
   1845\textrm{--}8325\ \textup{\AA}$) \hbox{wavelength-dependent} 
   lags generally follow $\tau(\lambda)\propto \lambda^{4/3}$, as 
   expected from a thin disk \citep{Cackett2020}; this result also
   supports the idea that the emission at $2500~\textup{\AA}$ likely
   comes from a thin disk.
   
 \subsection{Bolometric Luminosities and Eddington Ratios}
 \label{subsec:bol}
\begin{figure}
\centerline{
\includegraphics[scale=0.58]{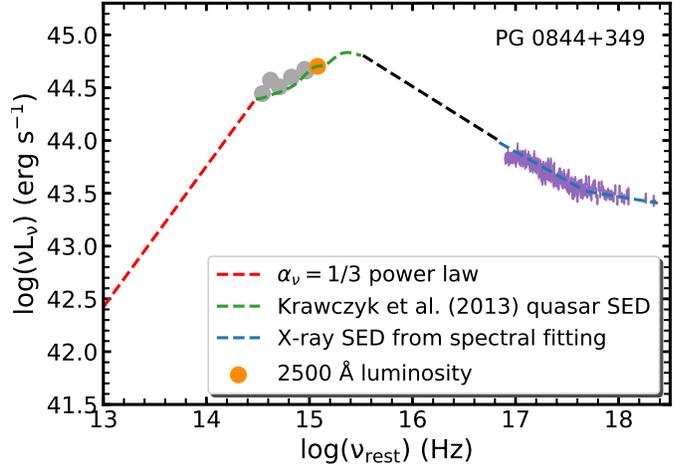}
}
\caption{An example of the SED used to estimate the bolometric 
luminosity. The gray dots show the observed UV/optical photometric data points (see Section~\ref{subsec:sed} for details). The purple symbols 
 represent the X-ray spectral data points that have been corrected for the Galactic absorption; 
the 5.5--7.5 keV spectrum was discarded due to strong iron K emission.
 The \hbox{IR-to-UV} SED template is normalized to the $2500~\textup{\AA}$ luminosity that is extrapolated from
the UV photometric data. The $912~\textup{\AA}\textrm{--}0.3$~keV SED (shown as the black dashed line) is a power law connecting the IR-to-UV SED template and the X-ray SED derived from spectral fitting (a \hbox{$0.3\textrm{--}2$ keV} {\sc zpowerlw} model plus 
a \hbox{$2\textrm{--}10$ keV} {\sc zpowerlw} model; see Tables~\ref{tbl2} and \ref{tbl3}).
 }
\label{fig-sed_mode}
\end{figure}

 Considering that most objects in our sample are nearby 
  moderate-luminosity AGNs, for which the \hbox{IR-to-UV} SEDs may 
  have significant contamination from the host galaxies, we    
  used an IR-to-UV quasar SED template scaled to the $2500~\textup{\AA}$ 
  luminosity plus the observed \hbox{X-ray} SED to estimate the 
bolometric luminosity for each object. An example of such an IR-to-X-ray SED (for PG $0844+349$)
is shown in Figure~\ref{fig-sed_mode}. We adopted the luminosity-dependent mean 
 quasar SED (low luminosity: ${\rm log}(\nu L_\nu)|_{\lambda=2500~\textup{\AA}} 
 \le 45.41$; mid luminosity: $45.41 \le {\rm log}(\nu L_\nu)|
 _{\lambda=2500~\textup{\AA}} \le 45.85$) in \cite{Krawczyk2013}  
 as the IR-to-UV template, and it was normalized to the $2500~\textup{\AA}$  
 luminosity that was derived from the UV/optical photometric data 
 (Sections~\ref{subsec:xmm}--\ref{subsec:swift}). As shown in 
 Section~\ref{subsec:sed} below, this mean quasar SED also describes 
 reasonably well the optical-to-UV SEDs of super-Eddington accreting AGNs.
 Most of the AGN IR ($\sim 1\textrm{--}30~\rm \mu m$) radiation is likely 
 reprocessed emission from the ``dusty torus'', which should not be included in 
 the computation of the bolometric luminosity \cite[e.g.,][and 
 references therein]{Krawczyk2013}. We thus replaced the 
 $1\textrm{--}30~\rm \mu m$ SED template with an $\alpha_{\nu}=1/3$
 power law \citep{Shakura1973} to account for 
 the IR emission from the accretion disk \citep[see discussion in Section~3.1 of][]{Davis2011}.
 
 In the \xray\ band, the rest-frame \hbox{2--10}~keV SED was obtained from the
 rest-frame $>2~\rm keV$ spectral fitting result (see 
 Section~\ref{subsec:spec}). 
  Soft excess emission is visible for most of our sample objects when extrapolating the hard X-ray \hbox{power-law} models to rest-frame $<2~\rm keV$ energies. In order to describe better  
  the shape of the soft X-ray SEDs, and thus measure the rest-frame 
  $0.3\textrm{--}2$ keV \xray\ luminosities, we applied simple models to 
  fit the soft X-ray spectra (\hbox{observed-frame} data from 0.3 keV to $2/(1+z)$~keV 
 for \xmm\ and \swift\ observations and 0.5 keV to 
$2/(1+z)$~keV for the \chandra\ observation). 
 The procedure is described as follows:
 (1) We first attempted to fit the soft \xray\ spectra 
 with a power-law model modified by Galactic absorption. 
 This model fits well the spectra for 24 objects. (2) For the other 23 objects, their soft \xray\ spectra 
are not described well by the power-law model with substantial residuals shown in the data vs.\ model plots. Therefore, we fitted  the soft \xray\ spectra with 
 a \hbox{thermal-Comptonization} component ({\sc comptt} in XSPEC) plus a power-law model, where the power-law component accounts for the coronal emission 
   and was fixed to that constrained from the rest-frame $>2~\rm keV$ 
   spectral fitting (Section~\ref{subsec:spec}). For five of the 23 objects (e.g., IRAS F$12397+3333$), 
   we also added an additional partial-covering ionized absorption component 
  ({\sc zxipcf}) into the fitting to account for weak absorption in the soft \hbox{X-rays};
  such absorption does not affect significantly the \hbox{rest-frame} $>2$ keV spectra.
 Table~\ref{tbl3} lists for each of our sample objects the best-fit 
 parameters in the soft \hbox{X-rays}. 
We note that the fitting method described above provides  phenomenological descriptions of the soft \xray\ continua for luminosity estimates. The best-fit models do not account for the broad-band X-ray spectra, and thus the best-fit parameters (e.g.,
 absorption column density, temperature of the \hbox{warm-corona} electrons) are not necessarily physical.  
  The final \hbox{rest-frame} 0.3--2~keV SEDs were derived 
  from the \hbox{best-fit} models that were corrected for Galactic  
absorption, and the rest-frame 0.3--2 keV luminosities are listed in Table~\ref{tbl3}. We set the $912~\textup{\AA}\textrm{--}0.3$~keV SED to a 
power law connecting the two endpoints of the \hbox{IR-to-UV} SED 
template and the X-ray SED \citep[e.g., Section 4.1 of][]{Laor1997}.

  Through integrating the constructed 
  \hbox{IR-to-X-ray} ($30~\mu \rm m\textrm{--}10~\rm keV$) SEDs, 
  we obtained the bolometric luminosities for our sample, 
  which span a range of $1.1\times10^{43}$ to 
$\rm 2.3\times 10^{46}~erg~s^{-1}$.
    We used the uncertainties on $L_{\rm 2500~\textup{\AA}}$ 
  and the 0.2--10 keV \xray\ luminosities to compute the measurement 
  uncertainties on $L_{\rm Bol}$.
  Given the $L_{\rm Bol}$ and $M_{\rm BH}$ values,
 we derived $L_{\rm Bol}/L_{\rm Edd}$ for our sample, 
 which range from $6.7\times10^{-3}$ to $5.5$, with a median 
 value of $0.14$. The uncertainties on 
  $L_{\rm Bol}/L_{\rm Edd}$ are propagated from the measurement 
  uncertainties on $L_{\rm Bol}$ and $M_{\rm BH}$. 
  The uncertainty on $L_{\rm Bol}/L_{\rm Edd}$ is 
  dominated by the $M_{\rm BH}$ uncertainty, and the contribution from 
  the $L_{\rm Bol}$ uncertainty is negligible. We note that there are 
  potential systematic uncertainties on both $M_{\rm BH}$ and $L_{\rm Bol}$ that may 
  introduce additional uncertainties for the $L_{\rm Bol}/L_{\rm Edd}$ 
  estimates (see discussion in Section~\ref{subsec:dis-gamma} below).
  Table~\ref{tbl1} lists the ${\rm log}(L_{\rm Bol}$) and ${\rm log}(L_{\rm Bol}/L_{\rm Edd}$) values and associated uncertainties.
    
\section{Results}
\label{sec:result}

\subsection{$\alpha_{\rm OX}$ versus $L_{2500~{\textup{\AA}}}$ Correlation}\label{subsec:aox-l2500}
\begin{figure}
\centerline{
\includegraphics[scale=0.58]{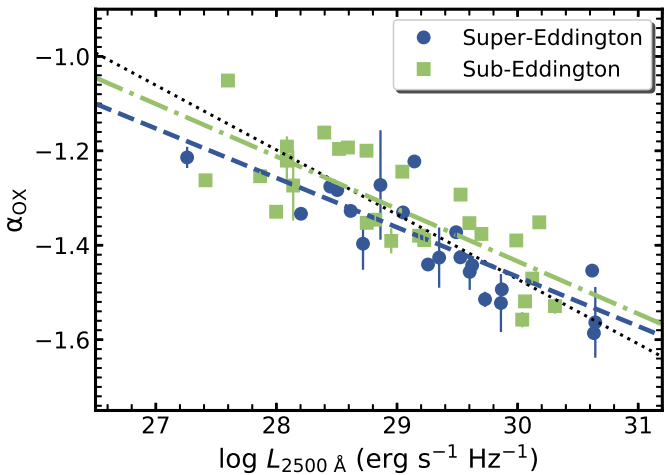}
}
\caption{X-ray-to-optical power-law slope ($\alpha_{\rm OX}$)
vs.~$2500~{\textup{\AA}}$ luminosity for
the \hbox{super-Eddington} subsample (blue circles) and the 
\hbox{sub-Eddington} subsample (green squares). 
The blue dashed (green dot-dashed) line shows the best-fit relation 
for the \hbox{super-Eddington} (\hbox{sub-Eddington}) subsample, given by Equation~\ref{eq2}
 (Equation \ref{eq3}). For comparison, the best-fit 
 relation of \cite{Steffen2006} is denoted with the black dotted
 line. }
\label{fig2}
\end{figure}

Figure~\ref{fig2} displays $\alpha_{\rm OX}$ versus 
${\rm log}(L_{2500~{\textup{\AA}}})$ for our \hbox{super-Eddington} and 
\hbox{sub-Eddington} subsamples. The two parameters are 
highly correlated for both subsamples. 
For the \hbox{super-Eddington} (\hbox{sub-Eddington}) subsample,
the Spearman rank correlation test gives a correlation coefficient 
 of $r_s=-0.84$ ($r_s=-0.77$) and a $p$-value of 
 $p=2.97\times 10^{-6}$ ($p=2.67\times 10^{-6}$). The $p$-value 
 indicates the probability of obtaining a correlation coefficient 
 $r_s$ at least as high as the observed one, under the null hypothesis
that the two sets of data are uncorrelated.

\begin{figure}
\centerline{
\includegraphics[scale=0.58]{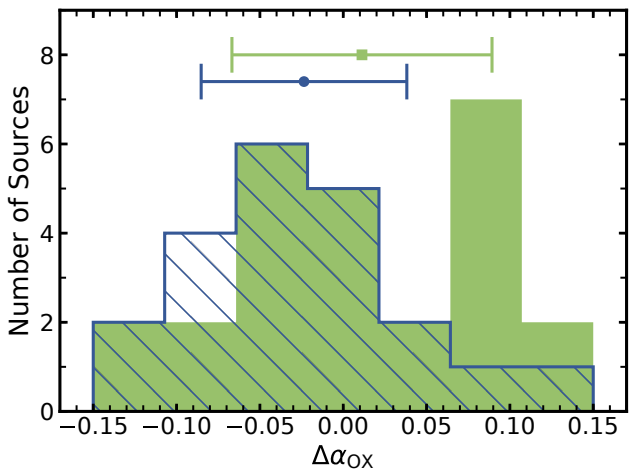}
}
\caption{Distribution of $\Delta\alpha_{\rm OX}$
for the super-Eddington (hatched blue histograms) and sub-Eddington
(filled green histograms) subsamples. $\Delta\alpha_{\rm OX}$ is 
defined as the difference between 
the observed $\alpha_{\rm OX}$ value and the one expected from the 
\cite{Steffen2006} relation (shown as the black dotted line in 
Figure~\ref{fig2}).
The two points with error bars
on the top show the mean values and $1\sigma$ deviations for the two 
subsamples. The two subsamples are consistent in the $\Delta\alpha_{\rm OX}$ distribution. 
}
\label{fig3}
\end{figure}
 
 We performed linear regression analysis on the $\alpha_{\rm OX}$ 
 versus ${\rm log}(L_{2500~{\textup{\AA}}})$ relations, using the 
 LINMIX\_ERR method \citep{Kelly2007}. This is a Bayesian method that 
 accounts for measurement uncertainties.
 For the \hbox{super-Eddington} subsample, the best-fit regression 
 equation with the $1\sigma$ uncertainty on each parameter is
\begin{equation}
\alpha_{\rm OX}=(-0.11\pm 0.02){\rm log}(L_{2500~{\textup{\AA}}})+(1.7\pm 0.6),
\label{eq2}
\end{equation}
with a scatter of 0.06. For the \hbox{sub-Eddington} subsample, the best-fit 
relation is  
\begin{equation}
\alpha_{\rm OX}=(-0.11\pm 0.02){\rm log}(L_{2500~{\textup{\AA}}})+(1.9\pm 0.5),
\label{eq3}
\end{equation} 
with a scatter of 0.08.
The relations (slopes and intercepts) for the two subsamples are consistent within their $1~\sigma$ uncertainties.

We then performed linear regression on the full sample, and
the best-fit relation is 
\begin{equation}
\alpha_{\rm OX}=(-0.11\pm 0.01){\rm log}(L_{2500~{\textup{\AA}}})+(1.9 \pm 0.4),
\label{eq1}
\end{equation} 
with a scatter of 0.07. The relation slope is consistent within 
the $1\sigma$ uncertainties with 
those for the \hbox{super-Eddington} and \hbox{sub-Eddington} 
subsamples. 
 Compared to previous results, our slope for the full sample is
  flatter than the slopes of $-0.14$ to $-0.22$ reported 
in most of the previous studies \citep{Strateva2005,Steffen2006,
Just2007,Gibson2008a,Lusso2010,Chiaraluce2018,Timlin2020}, but it is 
steeper than the slopes of $-0.06$ to $-0.07$ found by 
\cite{Green2009,Jin2012}.
\cite{Steffen2006} have suggested that the power-law 
slope of the $\alpha_{\rm OX}\textrm{--}L_{2500~{\textup{\AA}}}$ 
relation may be $L_{2500~{\textup{\AA}}}$ dependent, and it 
appears to be steeper towards higher $L_{2500~{\textup{\AA}}}$. 
Studies of this relation for high-luminosity quasars did find steeper 
slopes \cite[e.g.,][]{Gibson2008a,Timlin2020}. 
Therefore, the flat slope for our sample may be due to the generally 
lower UV luminosities. The best-fit relations for the two subsamples 
are plotted in Figure~\ref{fig2}.
For comparison, 
we also plotted in Figure~\ref{fig2} the relation of \cite{Steffen2006} with a dotted line.

 We further investigated the distribution of the 
$\Delta\alpha_{\rm OX}$ parameter, defined as the difference 
between the observed 
$\alpha_{\rm OX}$ value and the one expected from  
the $\alpha_{\rm OX}\textrm{--}L_{2500~{\textup{\AA}}}$ relation 
for typical AGNs.
Considering that our full sample may be biased toward  
super-Eddington accreting AGNs, we adopted the 
\cite{Steffen2006} $\alpha_{\rm OX}\textrm{--}L_{2500~{\textup{\AA}}}$ 
relation, $\alpha_{\rm OX}=-0.137\times {\rm log}(L_{2500~{\textup{\AA}}})+2.638$, 
which was derived from a large sample of 333 typical AGNs,
to compute the expected $\alpha_{\rm OX}$ values.
The parameter $\Delta\alpha_{\rm OX}$ is an indicator of 
the level of \xray\ weakness. As shown in Figure~\ref{fig3}, 
the $\Delta\alpha_{\rm OX}$ distributions for the
 two subsamples span the same range of $-0.15$ to 0.15, which is   
 within the rms scatter of the \cite{Steffen2006} 
 $\alpha_{\rm OX}\textrm{--}L_{2500~{\textup{\AA}}}$relation. 
  The mean and rms of the $\Delta\alpha_{\rm OX}$ values for the 
super-Eddington (sub-Eddington) subsample are $-0.024$ ($0.011$) and 
$0.062$ ($0.078$), respectively. Therefore, 
the \hbox{super-Eddington} subsample shows slightly weaker 
($\approx 23\%$ lower) \xray\ emission compared to the sub-Eddington 
subsample, but the significance of the difference is only $0.3\sigma$.
A \hbox{Kolmogorov-Smirnov} (KS) test on the  
 $\Delta\alpha_{\rm OX}$ distributions for the two subsamples yielded 
 $d=0.328$ and $p=0.131$, indicating that 
the two distributions are similar. We note that these results would 
not change significantly if we instead use the best-fit 
$\alpha_{\rm OX}\textrm{--}L_{2500~{\textup{\AA}}}$ 
relation (Equation~\ref{eq1}) for our full sample to
compute the expected $\alpha_{\rm OX}$ values.  
These results suggest that both the super- and 
 \hbox{sub-Eddington} subsamples show generally normal \xray\ 
 emission, when the high-state data are considered.

 \subsection{$L_{\rm 2~keV}$ versus $L_{2500~{\textup{\AA}}}$ Correlation} 
\begin{figure}
\centerline{
\includegraphics[scale=0.58]{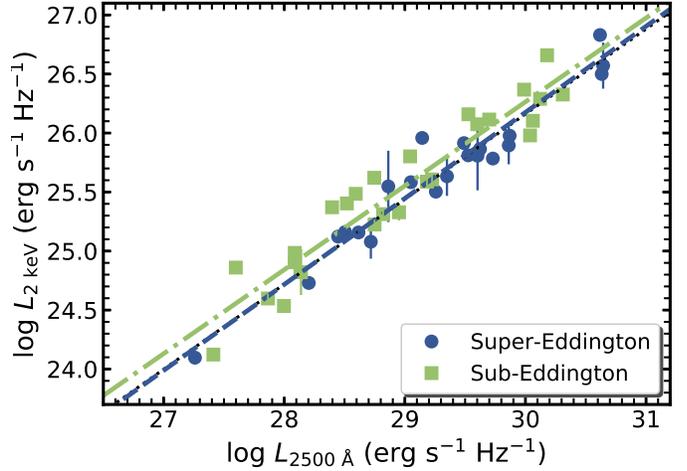}
}
\caption{Rest-frame 2~keV monochromatic luminosity 
vs.~\hbox{rest-frame} $2500~{\textup{\AA}}$ monochromatic luminosity 
for the \hbox{super-Eddington} subsample (blue circles) and the
 \hbox{sub-Eddington} subsample (green squares). 
 The blue dashed (green dot-dashed) line shows the \hbox{best-fit} 
 relation for the \hbox{super-Eddington} (\hbox{sub-Eddington}) 
 subsample, given by Equation~\ref{eq5} (Equation \ref{eq6}). 
 For comparison, the best-fit relation of \cite{Steffen2006} 
 is shown with the black doted line, which follows very closely with
 the blue-dashed line that denotes the relation for the 
 super-Eddington 
 subsample.}
\label{fig4}
\end{figure}

We also investigated the correlations between $L_{\rm 2~keV}$ and 
$L_{2500~{\textup{\AA}}}$ for our sample, shown in Figure~\ref{fig4}.
The correlations for both the 
super- and \hbox{sub-Eddington} subsamples are   
highly significant with Spearman coefficients of 
$r_s=0.91$ and $r_s=0.93$, respectively.  
We performed regression analysis with the LINMIX\_ERR method on 
the two sets of parameters. For the \hbox{super-Eddington} subsample, 
the best-fit relation is 
\begin{equation}
{\rm log}(L_{\rm 2~keV})=(0.73\pm0.05){\rm log}(L_{2500~{\textup{\AA}}})+(4.3\pm 1.4),
\label{eq5}
\end{equation} 
with a scatter of 0.17. 
For the \hbox{sub-Eddington} subsample, the best-fit relation is 
\begin{equation}
{\rm log}(L_{\rm 2~keV})=(0.71\pm 0.05){\rm log}(L_{2500~{\textup{\AA}}})+(5.0\pm 1.5),
\label{eq6}
\end{equation}
with a scatter of 0.22.   
The slopes and intercepts of the relations for the two subsamples
are consistent within the $1\sigma$ uncertainties.
The best-fit relation for the full sample is
\begin{equation}
{\rm log}(L_{\rm 2~keV})=(0.71\pm 0.03){\rm log}(L_{2500~{\textup{\AA}}})+(5.0 \pm 1.0) \ ,
\label{eq4}
\end{equation} 
with a scatter of 0.19. 
The slope of the relation for the full sample is consistent 
with those 
($\approx$ 0.7--0.8) reported
 in previous studies \citep[e.g.,][]{
Vignali2003,Strateva2005,Steffen2006,Just2007,Lusso2010,Lusso2016}.
We note that Equations~\ref{eq5}, \ref{eq6} and \ref{eq4}
 can be derived from Equations \ref{eq2}, \ref{eq3} and \ref{eq1}, 
 respectively, since $\alpha_{\rm OX}$ is 
 defined as the ratio between $L_{\rm 2~keV}$ and $L_{2500~{\textup{\AA}}}$.

\subsection{Correlation between $\Gamma$ and accretion rate}
\label{subsec:ga-mdot}
\begin{figure}
\centerline{
\includegraphics[scale=0.58]{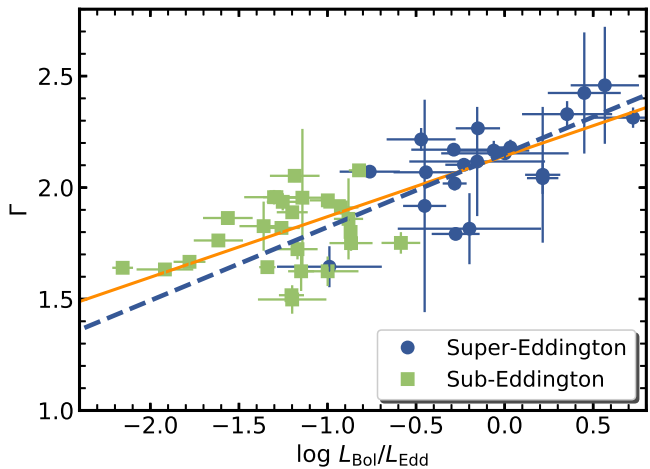}
}
\caption{Hard \xray\ photon index vs. $L_{\rm Bol}/L_{\rm Edd}$ for the \hbox{super-Eddington} subsample (blue circles) and 
the \hbox{sub-Eddington}
 subsample (green squares). The two subsamples overlap slightly 
 in the $L_{\rm Bol}/L_{\rm Edd}$ values, because they are classified 
 on the basis of $\dot{\rm \mathscr{M}}$ instead of $L_{\rm Bol}/L_{\rm Edd}$.
 The blue dashed (orange solid)  
 line shows the best-fit relation for the \hbox{super-Eddington} subsample (full sample), given by Equation~\ref{eq7} (Equation~\ref{eq9}).
 }
\label{fig-lledd}
\end{figure}

Figure~\ref{fig-lledd} plots $\Gamma$ versus ${\rm log}(L_{\rm Bol}/L_{\rm Edd})$ 
for our sample objects. For the full sample, a highly
significant correlation is present, and the Spearman rank 
correlation test resulted in a correlation coefficient of 
$r_s=0.72$ with a 
 \hbox{$p$-value} of $p=1.27\times 10^{-8}$. The correlation is  
 significant for the \hbox{super-Eddington} subsample ($r_s=0.60$ and 
 $p=0.004$); however, there is no statistically significant correlation 
 between $\Gamma$ and ${\rm log}(L_{\rm Bol}/L_{\rm Edd})$ for the 
 \hbox{sub-Eddington} subsample ($r_s=0.21$ and $p=0.30$). We then 
 performed linear regression analysis on the super-Eddington subsample and 
 the full sample with the LINMIX\_ERR method, 
 considering the measurement uncertainties on both $\Gamma$ and 
 $L_{\rm Bol}/L_{\rm Edd}$ in the fitting.  
 The \hbox{best-fit} relation for the 
 \hbox{super-Eddington} subsample is 
 \begin{equation} 
\Gamma=(0.33\pm 0.13){\rm log}(L_{\rm Bol}/L_{\rm Edd})+(2.15\pm 0.04),
 \label{eq7} 
 \end{equation} 
 with a scatter of 0.14. 
For the full sample, we obtained a best-fit relation: 
\begin{equation} 
\Gamma=(0.27\pm 0.04){\rm log}(L_{\rm Bol}/L_{\rm Edd})+(2.14\pm 0.04)\ ,
\label{eq9} 
\end{equation} with a scatter of 0.14. 
Our relation slope for the full sample is  
consistent with the slope ($0.31\pm 0.01$) reported in 
\cite{Shemmer2008} for their high-redshift quasars with BH masses 
determined based on the $\rm H\beta$ emission lines using the 
single-epoch virial mass 
 method. A similar slope ($0.32\pm 0.05$) was found by 
 \cite{Brightman2013} for their sample with BH masses obtained from 
 either $\rm H\beta$ or \ion{Mg}{2}-based single-epoch estimators.
 However, \cite{Risaliti2009} reported a steeper slope 
 ($0.58\pm 0.11$) for their subsample with $\rm H\beta$-based 
 single-epoch BH masses. A steep slope 
 ($\approx0.57$) was also found by \cite{Jin2012}, for their AGN 
sample with $\Gamma$ and $L_{\rm Bol}/L_{\rm Edd}$ estimated 
from the \hbox{UV/optical-to-X-ray} SED fitting.

\begin{figure}
\centerline{
\includegraphics[scale=0.58]{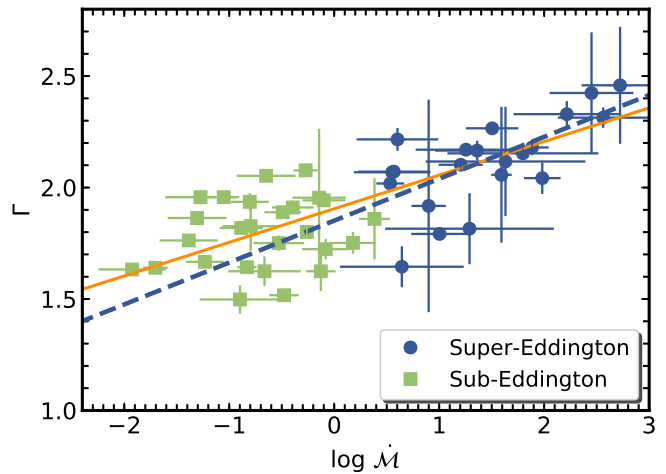}
}
\caption{Hard \xray\ photon index vs. normalized accretion rate 
for the \hbox{super-Eddington} subsample (blue circles) and 
the \hbox{sub-Eddington} subsample (green squares). 
The blue dashed (orange solid)  
 line shows the best-fit relation for the \hbox{super-Eddington} subsample (full sample), given by Equation~\ref{eq10} (Equation~\ref{eq12}).
 }
\label{fig-g-mdot}
\end{figure}

We also investigated the correlations between $\Gamma$ and   
normalized accretion rate ($\dot{\rm \mathscr{M}}$; see Figure~\ref{fig-g-mdot}). 
Similar to the trends between $\Gamma$ and $L_{\rm Bol}/L_{\rm Edd}$,
$\Gamma$ and $\dot{\rm\mathscr{M}}$ are highly 
correlated for the full sample ($r_s=0.70$ and $p=5.55\times 10^{-8}$) 
or the \hbox{super-Eddington} subsample ($r_s=0.63$ and 
$p=0.002$), but the correlation is not 
statistically significant for the \hbox{sub-Eddington} subsample  
($r_s=0.14$ and $p=0.493$). 
For the \hbox{super-Eddington} subsample, the \hbox{best-fit} 
relation is 
 \begin{equation} 
 \Gamma=(0.19\pm 0.08){\rm log}\dot{\rm \mathscr{M}}+(1.85\pm 0.11),
 \label{eq10} 
 \end{equation} 
 with a scatter of 0.14. 
For the full sample, the best-fit relation obtained 
from the linear regression analysis is 
\begin{equation}
\Gamma=(0.15 \pm 0.02){\rm log}\dot{\rm \mathscr{M}}+(1.91\pm 0.03).
\label{eq12}
\end{equation} 

The similar dependences of $\Gamma$ on  
$L_{\rm Bol}/L_{\rm Edd}$ and $\dot{\rm \mathscr{M}}$
indicate that $L_{\rm Bol}/L_{\rm Edd}$ and $\dot{\rm \mathscr{M}}$ 
are likely correlated. 
A Spearman rank correlation test indicates 
a highly significant correlation 
($r_s=0.97$ and $p=4.66\times 10^{-29}$) 
between $L_{\rm Bol}/L_{\rm Edd}$ and $\dot{\rm \mathscr{M}}$. 
Our linear regression analysis on log($L_{\rm Bol}/L_{\rm Edd}$) 
and $\rm log \dot{\rm \mathscr{M}}$ resulted in 
a slope of $0.53$.
Such a tight $L_{\rm Bol}/L_{\rm Edd} \textrm{--} \dot{\rm \mathscr{M}}$ 
relation has also been reported in \cite{Huang2020} for their quasar 
sample, where they  
found a power-law slope of $0.52$.
This relation can be naturally explained by the 
dependence of the two parameters on $M_{\rm BH}$: 
$\dot{\rm \mathscr{M}}$ mainly depends on
$M_{\rm BH}^{-2}$ (see Section~\ref{subsec:mdot}), 
and $L_{\rm Bol}/L_{\rm Edd}$ is proportional to 
$M_{\rm BH}^{-1}$. Therefore, $L_{\rm Bol}/L_{\rm Edd}$ and 
$\dot{\rm \mathscr{M}}$ 
are related in the form of $L_{\rm Bol}/L_{\rm Edd}\propto \dot{\rm 
\mathscr{M}}^{0.5}$, with some scatter associated with the 
distributions of $L_{\rm Bol}$ and $L_{2500~{\textup{\AA}}}$ for the 
sample AGNs. Given the similarities of the 
$\Gamma\textrm{--}L_{\rm Bol}/L_{\rm Edd}$ and 
$\Gamma\textrm{--}\dot{\rm \mathscr{M}}$ relations, 
we focus on the $\Gamma\textrm{--}L_{\rm Bol}/L_{\rm Edd}$ 
relations in the following discussion.

 \subsection{Spectral Energy Distributions of \hbox{Super-Eddington} 
 Accreting AGNs}
 \label{subsec:sed}
\begin{figure}
\centerline{
\includegraphics[scale=0.37]{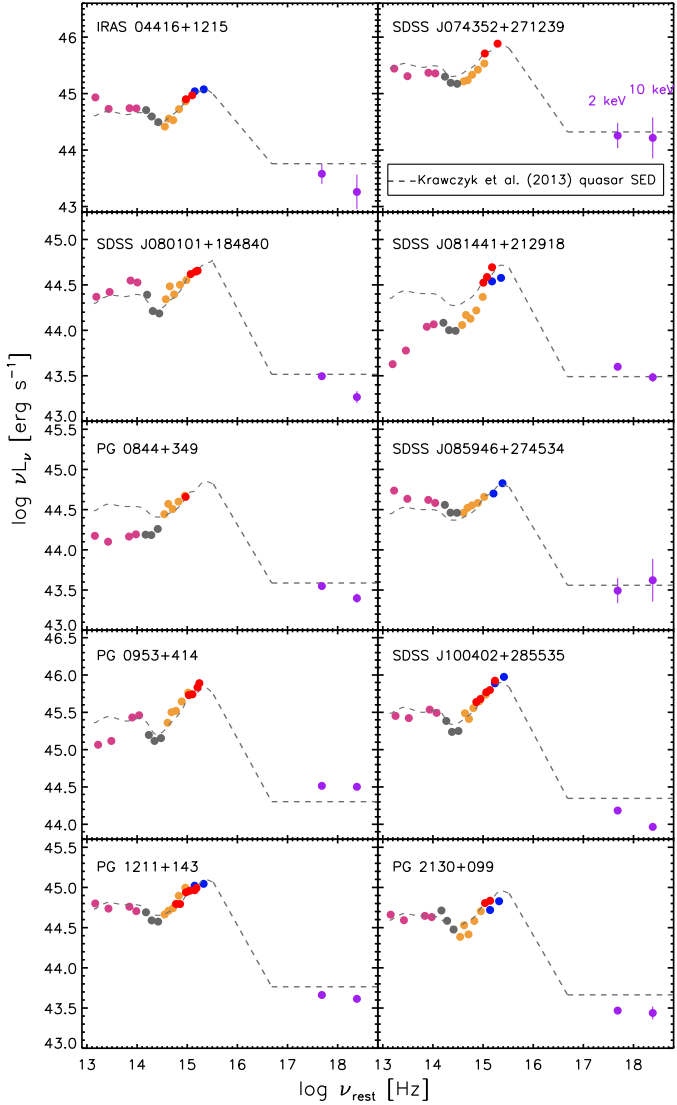}
}
\caption{IR-to-X-ray SEDs for the ten 
\hbox{super-Eddington} accreting AGNs with log$(\nu L_\nu)|_{\lambda=2500~\textup{\AA}} \ge 
 44.5$. 
The \hbox{IR-to-UV} photometric data points were gathered from
the {\it WISE} (magenta), 2MASS (gray), SDSS (orange), and {\it GALEX} 
(blue) catalogs.
The UVOT and OM photometric data are shown as red points, 
and the luminosities at 2~keV and 10~keV are shown as 
purple points. The dashed line in each panel shows the mean SED of
SDSS quasars \citep{Krawczyk2013}, which is
scaled to the mean $2500~{\textup{\AA}}$ luminosity extrapolated from
the UV photometric data of each object.
The SEDs may be affected by variability due to non-simultaneous 
multi-band observations (e.g., SDSS~J$074352+271239$ and 
SDSS~J$081441+212918$).
}
\label{fig-sed}
\end{figure}

 We constructed IR-to-X-ray SEDs for 
 ten \hbox{super-Eddington} accreting AGNs with UV luminosities 
 ($\nu L_\nu|_{\lambda=2500~\textup{\AA}}$) exceeding 
 $10^{44.5}\ \rm erg~s^{-1}$. This selection is based on the 
 consideration that the contamination from host galaxies should be 
 small in luminous AGNs. The photometric data were 
collected from the {\it Wide-field~Infrared~Survey~Explorer} 
\cite[{\it WISE};][]{Wright2010}, Two Micron All Sky Survey \cite[2MASS;]
[]{Skrutskie2006}, SDSS, and {\it GALEX} public catalogs.
 The UV/optical data were corrected for Galactic extinction 
 using the extinction law of \cite{Cardelli1989}. 
The SEDs are shown in Figure~\ref{fig-sed}.
We added the OM or UVOT data and the 2~keV and 10~keV 
luminosities.
 For comparison, the mean SED of typical SDSS quasars from 
 \cite{Krawczyk2013}, scaled to the $2500~\textup{\AA}$ luminosity 
 of each object, is plotted in each panel of Figure~\ref{fig-sed}.
 The IR-to-X-ray SEDs of most objects are  
 consistent with those of typical quasars, except for three objects
 (SDSS J$081441+212918$, PG $0844+349$, and PG~$0953+414$) that show 
 deficiencies in the mid-IR ({\it WISE}) bands. 

 The weak IR emission of PG~	$0844+349$ and PG~$0953+414$ has been 
 reported in \cite{Lyu2017}. PG~$0953+414$ was identified 
 as a \hbox{warm-dust-deficient} quasar, and PG $0844+349$ was an 
 ambiguous case but likely a \hbox{hot-dust-deficient} quasar.
The IR weakness of these three objects 
may be related to their \hbox{super-Eddington} accretion rates.
As suggested by \cite{Kawakatu2011}, \hbox{super-Eddington} 
accreting AGNs may tend to show weak IR emission due to the 
self-occultation effect
 of the thick accretion disk, which reduces the illumination of 
 the torus. However, these three IR weak objects do not show 
 extreme properties (e.g., $M_{\rm BH}$, $L_{\rm Bol}/L_{\rm Edd}$) 
 compared to the other seven objects, and they also exhibit typical 
 \hbox{optical-to-X-ray} SEDs. PG $0844+349$ has been found
to show extreme X-ray variability \citep[e.g.,][]{Gallagher2001,Gallo2011}, but the 
other extremely X-ray variable AGN among these ten objects, 
PG $1211+143$, does not have IR deficiency. We thus do not find any 
distinctive feature that may be related to the IR deficiency. Further 
investigations, probably utilizing a larger SED sample, are required to 
confirm and understand the potential IR deficiency of \hbox{super-Eddington} accreting AGNs.

 \section{Discussion} 
 \label{sec:discuss}

 \subsection{X-ray Emission Strength of \hbox{Super-Eddington} Accreting AGNs}
 \label{subsec:dis-aox}
 We examined the correlations between 
 $\alpha_{\rm OX}$ and $L_{2500~{\textup{\AA}}}$
 for the super- and \hbox{sub-Eddington} subsamples, in
order to determine whether \hbox{super-Eddington} accreting AGNs show 
different \xray\ emission strength relative to UV/optical emission,
 compared to \hbox{sub-Eddington} accreting AGNs. 
Significant $\alpha_{\rm OX}\textrm{--}L_{2500~{\textup{\AA}}}$ 
correlations were confirmed for both subsamples.
Compared to the \hbox{sub-Eddington} subsample, 
 the best-fit relation between $\alpha_{\rm OX}$ 
  and ${\rm log}(L_{2500~{\textup{\AA}}})$ for the 
  \hbox{super-Eddington} 
 subsample has a slightly flatter slope and a smaller intercept, but 
 the parameters are consistent considering the $1\sigma$ 
 uncertainties. 
The two subsamples also show   
 similar $\Delta\alpha_{\rm OX}$ distributions (see Figure~\ref{fig3}), with the ranges within 
 the rms scatter of the \cite{Steffen2006} 
 $\alpha_{\rm OX}\textrm{--}L_{2500~{\textup{\AA}}}$ relation. 
 These results suggest that \hbox{super-Eddington} 
 accreting AGNs show normal \xray\ emission strength and 
 follow a similar $\alpha_{\rm OX}\textrm{--}L_{2500~{\textup{\AA}}}$ 
 relation as \hbox{sub-Eddington} accreting AGNs or typical AGNs,
 when their high-state \xray\ data are considered.

A few studies of IMBH candidates with high Eddington
ratios have revealed that a large fraction of IMBHs 
 deviate significantly (with $\Delta\alpha_{\rm OX} \la -0.25$; corresponding to $\ga 2\sigma$ deviations) 
 from the $\alpha_{\rm OX}\textrm{--}L_{2500~{\textup{\AA}}}$ 
 relation for typical AGNs \citep[e.g.,][]{Greene2007,Dong2012}.
 We consider that the \xray\ weakness of these IMBHs may be caused by 
\xray\ absorption, as some objects show unusually flat 
\xray\ spectra. Our investigation shows 
 that \hbox{super-Eddington} accreting AGNs tend to show strong \xray\ 
variability, likely related to shielding by the thick 
 accretion disk and/or its associated outflow in the low states 
 (see discussion in Section~\ref{subsec:dis-xvar} bellow). In this 
 study, we have intentionally selected \hbox{high-state} 
 observational data to probe the intrinsic \xray\  
 properties of our sample. 
 Mixing high- and low-state data 
 could reveal a fraction of objects deviating from the expected 
 $\alpha_{\rm OX}\textrm{--}L_{2500~{\textup{\AA}}}$ relation,
 showing different levels of \xray\ weakness.

Equivalent to the $\alpha_{\rm OX}\textrm{--}L_{2500~{\textup{\AA}}}$ 
correlations, strong and consistent correlations between 
$L_{\rm 2~keV}$ and $L_{2500~{\textup{\AA}}}$ for the 
super- and \hbox{sub-Eddington} subsamples were also found. 
Although the   
$L_{\rm 2~keV}\textrm{--}L_{2500~{\textup{\AA}}}$ and $\alpha_{\rm OX}\textrm{--}L_{2500~{\textup{\AA}}}$ relations are 
strong, the scatters of the two relations are large, as also noted
in previous studies \citep[e.g.,][]
{Vignali2003,Strateva2005,Steffen2006,Just2007,Lusso2010,Lusso2016}.
The scatters may be caused by factors such as measurement 
uncertainties, \xray\ absorption, host-galaxy contamination, 
and intrinsic scatter related to differences in AGN physical 
properties \citep[e.g.,][]{Lusso2016}. 
There are only slight ($< 1\sigma$) 
differences between the power-law slopes and intercepts of the 
$\alpha_{\rm OX}\textrm{--}L_{2500~{\textup{\AA}}}$ 
($L_{\rm 2~keV}\textrm{--}L_{2500~{\textup{\AA}}}$)
relation for the two subsamples, suggesting that 
the different accretion physics in super- and 
\hbox{sub-Eddington} accreting AGNs likely contributes little to 
the intrinsic scatter of these relations.

The $\alpha_{\rm OX}\textrm{--}L_{2500~{\textup{\AA}}}$ or
$L_{\rm 2~keV}\textrm{--}L_{2500~{\textup{\AA}}}$
relation indicates that the fraction of accretion-disk radiation (or 
equivalently the accretion power in the radiatively-efficient case) 
dissipated via the corona has a strong dependence on the UV 
luminosity. More optically luminous AGNs are observed to produce 
relatively weaker \xray\ emission from their coronae. 
There is still no clear understanding of the physics behind this 
empirical disk-corona connection. Simple qualitative explanations   
usually involve how the accretion power dissipation formula or the 
coronal size/structure changes with accretion rate 
(e.g., \citealt{Merloni2003,Wang2004-ratio,Yang2007,Lusso2017,
Kubota2018,Wang2019,Jiang2019,Arcodia2019}; see more discussion 
in Section~\ref{subsec:aox-lledd}). Nevertheless, our finding here 
that super-Eddington accreting AGNs follow basically the same 
$\alpha_{\rm OX}\textrm{--}L_{2500~{\textup{\AA}}}$  
($L_{\rm 2~keV}\textrm{--}L_{2500~{\textup{\AA}}}$) relation 
as that for typical AGNs suggests that super-Eddington accreting 
AGNs, regardless of their geometrically thick accretion disks and 
the potential photon-trapping effects, likely share 
 the same relation when dissipating the accretion power 
between the accretion disk and corona as sub-Eddington accreting AGNs.
Alternatively, our finding may suggest that 
\hbox{super-Eddington} accreting AGNs, or at least the AGNs in our 
\hbox{super-Eddington} subsample, probably 
do not have distinctive accretion physics (e.g., no thick disks)
compared to sub-Eddington accreting AGNs.

Our finding indicates that typical AGNs with ${\rm log}(L_{2500~{\textup{\AA}}})$ in the range of $\sim 27.3\textrm{--}30.6\ \rm (erg\ s^{-1}\ Hz^{-1})$ all follow the same 
 $\alpha_{\rm OX}\textrm{--}L_{2500~{\textup{\AA}}}$ relation.
 After accounting for the scatter, this relation may indeed be used to estimate the intrinsic \xray\ luminosity for an AGN/quasar given its UV/optical luminosity, and then to identify \xray\ weak AGNs (e.g., \citealt{Gibson2008a,Pu2020}) or to measure enhanced \xray\ emission in radio-loud AGNs \citep[e.g.,][]{Miller2011,Zhu2020}. 

 \begin{figure*}
\centerline{
\includegraphics[scale=0.58]{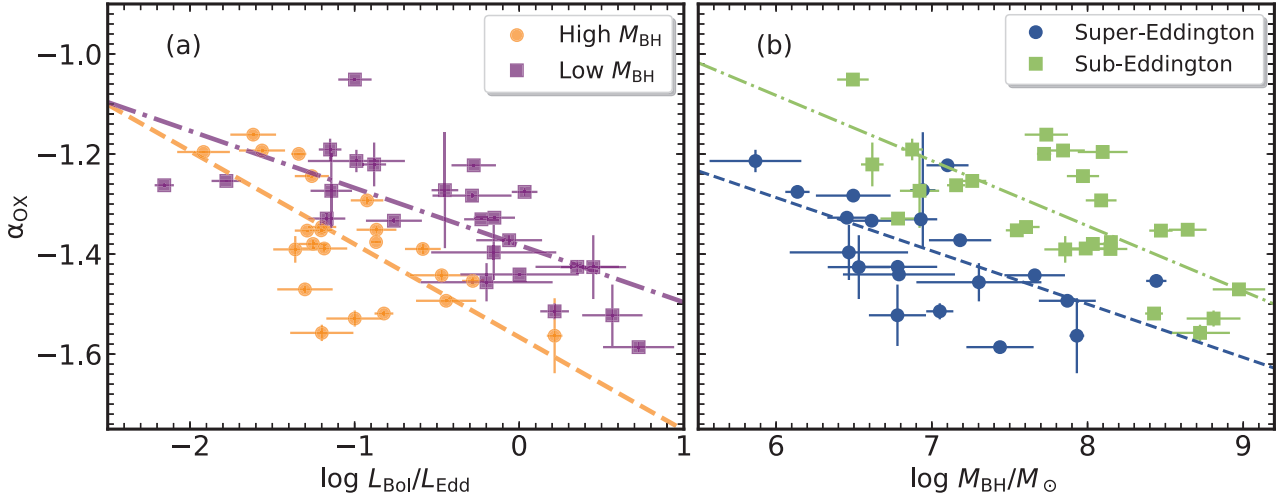}
}
\caption{X-ray-to-optical power-law slope ($\alpha_{\rm OX}$) vs. 
(a) $L_{\rm Bol}/L_{\rm Edd}$ and (b) $M_{\rm BH}$.
 In Panel (a), the orange circles represent 
high-$M_{\rm BH}$ objects with BH masses larger than the median value 
($10^{7.44}~M_\odot$) of the full sample, 
and the purple squares represent the low-$M_{\rm BH}$ objects.
The best-fit relations for the high-$M_{\rm BH}$ objects (orange dashed line) and low-$M_{\rm BH}$ objects (purple dot-dashed line) are plotted to guide the eye.
In Panel (b), the \hbox{super-Eddington} (\hbox{sub-Eddington}) subsample is shown as blue circles (green squares), with the best-fit relation shown as the blue dashed (green dot-dashed) line.
   }
\label{aox-lledd}
\end{figure*}

\subsection{A More Fundamental $\alpha_{\rm OX}$ versus
$L_{\rm Bol}/L_{\rm Edd}$ plus $M_{\rm BH}$ Relation?}
\label{subsec:aox-lledd}  
Our carefully constructed sample of AGNs with the best available 
BH-mass measurements provides a good opportunity for seeking a 
physical explanation of the observed 
$\alpha_{\rm OX}\textrm{--}L_{2500~{\textup{\AA}}}$ relation.
In this section, we explore the possibility that the 
$\alpha_{\rm OX}\textrm{--}L_{2500~{\textup{\AA}}}$ relation is physically  
 driven by the dependences of $\alpha_{\rm OX}$ on the two fundamental 
 parameters, $L_{\rm Bol}/L_{\rm Edd}$ and $M_{\rm BH}$ 
 \citep[e.g.,][]{Shemmer2008}.
 
 One promising explanation 
 for the formation of the corona is that the magnetic field 
 amplified by the magneto-rotational instability (MRI) saturates due
  to vertical buoyancy, and it extends outside the accretion 
 disk and forms a magnetically dominated coronal region 
 \citep[e.g.,][]{Stella1984,Tout1992,Svensson1994,Miller2000,
 Merloni2002,Blackman2009,Jiang2014}.
 A fraction ($f_{\rm X}$) of the accretion power carried away by the magnetic 
 buoyancy is released via magnetic reconnection, thereby heating the 
 corona \cite[e.g.,][]{Galeev1979,DiMatteo1998,Liu2002,Uzdensky2008}.
Based on these descriptions and the basic theory of the standard 
accretion disk \citep{Shakura1973}, 
analytic models of the accretion disk-corona system
\citep[e.g.,][]{Merloni2003,
 Wang2004-ratio,Yang2007,Cao2009,Lusso2017,
 Kubota2018,Wang2019,Arcodia2019,Cheng2020}
 predict a smaller energy dissipation fraction $f_{\rm X}$ for 
 an accretion disk with a higher $L_{\rm Bol}/L_{\rm Edd}$,
as the accretion disk becomes more radiation-pressure dominated and 
the MRI grows less rapidly. Besides, the radiation 
magneto-hydrodynamic (MHD) simulations by \cite{Jiang2014,Jiang2019}
also suggest a weaker (smaller $f_{\rm X}$) and more compact corona 
when the accretion rate increases. 
 A similar, albeit weaker, trend between $f_{\rm X}$ and $M_{\rm BH}$
 is also expected \citep[e.g., Figure~5 of][]{Yang2007}, as the gas 
 pressure decreases more rapidly than the radiation pressure when  
 $M_{\rm BH}$ increases and the disk is again more radiation-pressure 
 dominated with relatively weaker MRI.

The parameter $\alpha_{\rm OX}$, as an indicator of the coronal 
\xray\ emission strength relative to accretion-disk 
UV/optical emission, is likely dependent on 
 the fraction of accretion energy released in the corona. 
 As discussed above, an AGN with higher $L_{\rm Bol}/L_{\rm Edd}$ 
 and/or $M_{\rm BH}$ has a smaller $f_{\rm X}$, and thus the corona is 
 relatively weaker, leading to a smaller (steeper) $\alpha_{\rm OX}$. 
 Therefore, $\alpha_{\rm OX}$ is expected to be inversely correlated 
 with $L_{\rm Bol}/L_{\rm Edd}$ or $M_{\rm BH}$ when the other 
 parameter is fixed.  
We thus investigated whether these expected correlations 
exist for our sample. 
It is shown in Figure~\ref{aox-lledd}(a) that 
 $\alpha_{\rm OX}$ is \hbox{anti-correlated} 
 with $L_{\rm Bol}/L_{\rm Edd}$, 
 and the correlation appears more significant when breaking the full
 sample into the high-$M_{\rm BH}$ and low-$M_{\rm BH}$ subsamples.
Moreover, at a fixed $L_{\rm Bol}/L_{\rm Edd}$, objects with higher 
 $M_{\rm BH}$ systematically have lower $\alpha_{\rm OX}$, 
 which implies a dependence of $\alpha_{\rm OX}$ on $M_{\rm BH}$.
 Such an \hbox{anti-correlation} does exist, as shown in 
 Figure~\ref{aox-lledd}(b). The correlation appears more 
 significant when limiting to the \hbox{super-Eddington} 
 or sub-Eddington subsample. 
 We performed partial correlation analysis using the R package 
 {\it ppcor} \citep{Kim2015} on $\alpha_{\rm OX}$ versus 
 $L_{\rm Bol}/L_{\rm Edd}$ ($M_{\rm BH}$), 
 controlling for $M_{\rm BH}$ ($L_{\rm Bol}/L_{\rm Edd}$).
 The $\alpha_{\rm OX}\textrm{--}L_{\rm Bol}/L_{\rm Edd}$ 
 ($\alpha_{\rm OX}\textrm{--}M_{\rm BH}$) correlation is highly significant when controlling for $M_{\rm BH}$ ($L_{\rm Bol}/L_{\rm Edd}$), with a Spearman correlation coefficient of $-0.74$ ($-0.69$) and a $p$-value of $3.00\times 10^{-9}$ ($1.11\times 10^{-7}$). 
 The dependence of $\alpha_{\rm OX}$ on $L_{\rm Bol}/L_{\rm Edd}$ or $M_{\rm BH}$ has been discussed in previous studies. 
 Some authors found a significant correlation between $\alpha_{\rm OX}
 $ and $L_{\rm Bol}/L_{\rm Edd}$ \citep[e.g.,][]
 {Shemmer2008,Grupe2010,Lusso2010,Wujian2012,Jin2012,Chiaraluce2018}, 
 while some found no significant correlation 
 \citep[e.g.,][]{Vasudevan2007,Done2012}. 
 Some authors found a significant correlation between 
 $\alpha_{\rm OX}$ and $M_{\rm BH}$ \citep[e.g.,][]{Done2012,Chiaraluce2018}. 
Our results above suggest that $\alpha_{\rm OX}$ likely depends on 
 both $L_{\rm Bol}/L_{\rm Edd}$ and $M_{\rm BH}$. Thus the scatter of 
 the correlation with solely $L_{\rm Bol}/L_{\rm Edd}$ or $M_{\rm BH}$ is considerable.
 
  \begin{figure*}
\centerline{
\includegraphics[scale=0.58]{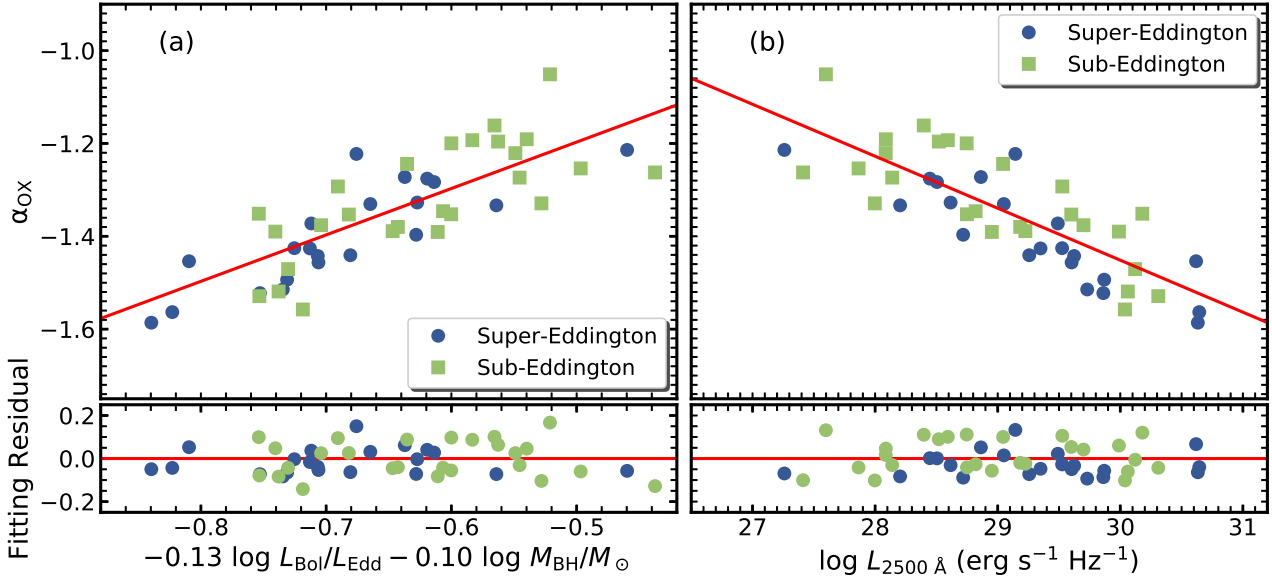}
}
\caption{(a) $\alpha_{\rm OX}\textrm{--} L_{\rm Bol}/L_{\rm Edd} \textrm{--}M_{\rm BH}$ 
plane seen edge-on. The red solid line shows the best-fit relation 
given by Equation~\ref{eq_aox_ledd}. (b) $\alpha_{\rm OX}$ 
vs. $2500\textup{\AA}$ luminosity. The red solid 
line shows the best-fit relation given by Equation~\ref{eq1}. The 
bottom panels show the fitting residuals, defined as the differences
between the observed $\alpha_{\rm OX}$ values and the expectations 
from the corresponding best-fit relation. 
The two relations show comparable scatters. 
}
\label{aox-mode}
\end{figure*}

 We performed multi-variate linear regression on the relation 
 $\alpha_{\rm OX}=\beta {\rm log}(L_{\rm Bol}/L_{\rm Edd})+\gamma{\rm log}M_{\rm BH}+\delta$ using
 the Python package {\it emcee} \citep{Foreman2013}, which is a Python 
 implementation of Goodman \& Weare's \hbox{affine-invariant} 
 Markov chain Monte Carlo (MCMC) ensemble sampler.
 The measurement uncertainties on the three parameters are included in 
 the fitting. The best-fit relation is
\begin{align}
\alpha_{\rm OX}=(-0.13 \pm 0.01){\rm log}(L_{\rm Bol}/L_{\rm Edd})-  \nonumber\\
(0.10 \pm 0.01){\rm log}M_{\rm BH}-(0.69\pm 0.09)
\label{eq_aox_ledd}
\end{align}
with a scatter of 0.07.
 An edge-on view of the $\alpha_{\rm OX}\textrm{--} L_{\rm Bol}/L_{\rm Edd} \textrm{--}M_{\rm BH}$ 
three-dimensional plane is shown in Figure~\ref{aox-mode}(a). 
The $\alpha_{\rm OX}\textrm{--} L_{\rm Bol}/L_{\rm Edd} \textrm{--}M_{\rm BH}$ relation may be the physical origin of the observed 
$\alpha_{\rm OX}\textrm{--} L_{2500\textup{\AA}}$ relation, 
as $L_{\rm Bol}/L_{\rm Edd}\times M_{\rm BH}\propto L_{\rm Bol}\propto 
L_{2500\textup{\AA}}$.
For comparison, we plotted the $\alpha_{\rm OX}$ versus ${\rm log}(L_{\rm 2500~\textup{\AA}})$ relation 
for our full sample in Figure~\ref{aox-mode}(b). 
We note that the scatter of the $\alpha_{\rm OX}\textrm{--} 
L_{\rm Bol}/L_{\rm Edd} \textrm{--}M_{\rm BH}$ relation 
is comparable to that 
of the $\alpha_{\rm OX}\textrm{--} L_{2500\textup{\AA}}$ relation, 
which is probably due to the large uncertainties on both 
$L_{\rm Bol}/L_{\rm Edd}$ and $M_{\rm BH}$. 
It could also be due to the dependence of 
$\alpha_{\rm OX}$ on a third parameter, the ratio of the gas plus radiation pressure to the 
magnetic pressure, as this ratio may work together with $L_{\rm Bol}/L_{\rm Edd}$ and $M_{\rm BH}$ to
determine the broad-band AGN SED \citep[e.g.,][]{Cheng2020}.

We note that it is difficult to determine if 
the relation between $\alpha_{\rm OX}$ 
and $L_{\rm Bol}/L_{\rm Edd}$ plus $M_{\rm BH}$ is more fundamental
 than the $\alpha_{\rm OX}\textrm{--}L_{\rm 2500~\textup{\AA}}$ 
 relation, or if it is a secondary manifestation of the observed 
 $\alpha_{\rm OX}\textrm{--}L_{\rm 2500~\textup{\AA}}$ relation.
There is a tight linear correlation ($r_s\approx 1$) between 
 $L_{\rm Bol}$ ($\propto L_{\rm Bol}/L_{\rm Edd} \times M_{\rm BH}$) and 
 $L_{\rm 2500~\textup{\AA}}$ for our sample objects. Therefore, from   
the $\alpha_{\rm OX}\textrm{--}L_{\rm 2500~\textup{\AA}}$ 
 relation (equivalently an $\alpha_{\rm OX}\textrm{--}L_{\rm Bol}$ relation),
a significant partial correlation between 
$\alpha_{\rm OX}$ and 
$L_{\rm Bol}/L_{\rm Edd}$ or $M_{\rm BH}$ when controlling for the other parameter is expected.
 We performed a test through creating 
 mock sets of parameter $K$ to replace $M_{\rm BH}$.
 In each realization, the $K$ values are randomly distributed 
in the same range as that of $M_{\rm BH}$ for our sample, 
and we then analyzed the 
partial correlation between $\alpha_{\rm OX}$ and $L_{\rm 2500~\textup{\AA}}/K$ when controlling for $K$.
A number of realizations with different $K$ values generated  
correlation coefficients of $-0.6$ to $-0.8$ and 
\hbox{$p$-values} of $10^{-7}\textrm{--}10^{-9}$. These correlation 
significance levels are similar to those of $\alpha_{\rm OX}$ against 
$L_{\rm Bol}/L_{\rm Edd}$ or $M_{\rm BH}$.
Plots of $\alpha_{\rm OX}$ versus $L_{\rm 2500~\textup{\AA}}/K$ are
also similar to the $\alpha_{\rm OX}\textrm{--}L_{\rm Bol}/L_{\rm Edd}
$ relation presented in Figure~\ref{aox-lledd} (a).
Therefore, we cannot determine whether the $\alpha_{\rm OX}$ versus 
$L_{\rm Bol}/L_{\rm Edd}$ plus $M_{\rm BH}$ correlation is 
fundamental, although physically this is an appealing explanation.  
A possible method to resolve this issue is to 
investigate these correlations for individual AGNs, such as \hbox{changing-look} AGNs varying in accretion rate.  
Without the complication from $M_{\rm BH}$,
we may constrain better  
the dependence of $\alpha_{\rm OX}$ on $L_{\rm Bol}/L_{\rm Edd}$.

\subsection{Relation between $\Gamma$ and $L_{\rm Bol}/L_{\rm Edd}$}
\label{subsec:dis-gamma}
A strong correlation between $\Gamma$ and $L_{\rm Bol}/L_{\rm Edd}$ 
($\dot{\rm \mathscr{M}}$) was confirmed for our full sample and \hbox{super-Eddington} 
subsample. However, such a correlation is not statistically significant for the 
\hbox{sub-Eddington} subsample. We caution that large uncertainties 
associated with the measurements of $L_{\rm Bol}/L_{\rm Edd}$ and $\Gamma$
may introduce considerable uncertainties for 
the $\Gamma\textrm{--}{\rm log}(L_{\rm Bol}/L_{\rm Edd})$ relation. 
The BH masses of our sample were obtained from the RM method, 
 which is arguably the most reliable method for AGN \hbox{BH-mass} 
 measurements. However, the RM method is based on the assumption of 
 virial gas motions in the 
 broad-line regions (BLRs), which may not be valid for 
 \hbox{super-Eddington} accreting systems due to the impact of 
 the large radiation pressure and the anisotropy of the ionizing radiation \citep[e.g.,][] 
 {Marconi2008,Marconi2009,Netzer2010,Krause2011,Pancoast2014,Li2018}.
In addition, there are potential uncertainties associated 
with the measurements of $L_{\rm Bol}$. The approach we used to
 obtain $L_{\rm Bol}$ is similar to the estimation through the use of
 bolometric corrections, which employ approximations to the mean 
 properties of typical quasars. The main improvement in our study 
 is that the \xray\ spectral shapes for individual objects were taken
  into account. Additional uncertainties on $L_{\rm Bol}$ may come 
  from the \hbox{UV-to-X-ray SED} which was set to a simple 
 power law. Super-Eddington accreting AGNs are expected to emit 
 excess EUV radiation compared to sub-Eddington accreting AGNs 
 \cite[e.g.,][]{Castell2016,Kubota2018}, although there is still no 
 clear observational evidence due to the lack of EUV data. 
Moreover, the criterion of $L_{\rm Bol}/L_{\rm Edd}$ (or 
$\dot{\rm \mathscr{M}}$) for identifying \hbox{super-Eddington} 
accreting AGNs is also rather uncertain 
\citep[e.g.,][]{Laor1989,Sadowski2011,Sadowski2016}.

The choice of \xray\ energy band in spectral fitting may also 
introduce uncertainties on the derived photon indices. 
The \xray\ energy band investigated in this study is the rest-frame 
$>\rm 2~keV$ band. This choice is based on the idea that the \xray\ 
spectrum in this band is generally insensitive to contamination
 from the potential soft-excess component or absorption 
 \citep[e.g.,][]{Shemmer2006,Shemmer2008,Risaliti2009,Brightman2013}.
 However, although there is no clear evidence of 
 soft excesses and absorption in the hard \xray\ spectra of 
 our sample AGNs, their influence might not be completely 
 eliminated.  
 

   With the above caveats in mind, we do find a significant $\Gamma\textrm{--}L_{\rm Bol}/L_{\rm Edd}$ relation for the full sample. 
   Compared to similar relations reported in previous 
  studies (see Section~\ref{subsec:ga-mdot}), there are notable 
  discrepancies in the \hbox{power-law} slopes.
  These discrepancies might be due to the different samples, 
different $M_{\rm BH}$ ($L_{\rm Bol}/L_{\rm Edd}$) estimation methods,
  or different statistical methods used in these studies (also see Section~4.3 of \citealt{Brandt2015}). 
 A large unbiased sample with reliable parameter 
 measurements and covering a wide range of $L_{\rm Bol}/L_{\rm Edd}$ 
 is required to establish the $\Gamma\textrm{--}L_{\rm Bol}/L_{\rm Edd}$ relation for general AGNs.
 The \xray\ photon index could then, if confirmed, serve as 
 an \hbox{Eddington-ratio} indicator, provided that the large scatter of the $\Gamma\textrm{--}L_{\rm Bol}/L_{\rm Edd}$ relation is understood and taken 
 into account.

There is not a significant $\Gamma\textrm{--}L_{\rm Bol}/L_{\rm Edd}$  correlation for the sub-Eddington subsample. 
 The best-fit relation for the super-Eddington subsample does not 
 appear to differ significantly from that for the full sample either. 
 Therefore, we cannot constrain any difference between the super- and
  \hbox{sub-Eddington} accreting AGNs
 in terms of the $\Gamma\textrm{--}L_{\rm Bol}/L_{\rm Edd}$ 
 relation. However, a few recent studies 
  have reported that \hbox{super-Eddington} accreting AGNs have 
  even steeper \xray\ photon indices in excess of those expected from 
  the $\Gamma\textrm{--}L_{\rm Bol}/L_{\rm Edd}$ relation for 
 \hbox{sub-Eddington} accreting AGNs \citep[e.g.,][]
 {Gliozzi2020,Huang2020}. If such a finding is 
 real, any physical explanations must involve the expected properties 
 of \hbox{super-Eddington} accretion disks, while maintaining 
 basically the same accretion power dissipation relation for the 
 accretion disk-corona system 
 (see the discussion in Section~\ref{subsec:dis-aox}). 
 A possible physical explanation is that 
in the thick disk of a \hbox{super-Eddington} AGN, more radiation is
emitted from the inner region of the disk due to the longer diffusion 
timescale for photons to escape from the disk surface and the stronger 
magnetic buoyancy in the inner region \cite[e.g.,][]{Jiang2014}; 
this effect increases the UV/optical emission received by the 
compact corona, reduces its temperature and optical depth, and leads to 
an even steeper \xray\ spectrum. Nevertheless, a larger sample of 
sub-Eddington accreting AGNs with RM measurements is required to allow 
further investigations of any difference between the 
$\Gamma\textrm{--}L_{\rm Bol}/L_{\rm Edd}$ correlations for super- 
and sub-Eddington accreting AGNs.

 \subsection{The Impact of X-ray Variability}
\label{subsec:dis-xvar}
 Our study suggests that \hbox{super-Eddington} accreting AGNs exhibit
 normal \xray\ emission and generally follow  
 the same $\alpha_{\rm OX}\textrm{--}L_{2500~{\textup{\AA}}}$ ($L_{\rm 2~keV}\textrm{--}L_{2500~{\textup{\AA}}}$)
relation as \hbox{sub-Eddington} accreting AGNs, as long as their intrinsic 
\xray\ emission is considered.
  However, a fraction of \hbox{super-Eddington} accreting AGNs 
  tend to show extreme large-amplitude (factors of $>10$) \xray\ variability 
 (e.g., 1H~$0707-495$: \citealt{Fabian2012}; IRAS~$13224-3809$: \citealt{Boller1997,Jiang2018}; SDSS J$075101.42+291419.1$: \citealt{Liu2019}).
Three \hbox{super-Eddington} accreting AGNs (Mrk~335, PG~$1211+143$, and PG~$0844+349$) in our sample have 
varied in \xray\ flux by factors of larger than 10, while 
they have not shown coordinated UV/optical continuum or 
\hbox{emission-line} variability \citep[e.g.,][]
{Guainazzi1998,Peterson2000,Grupe2007a,Bachev2009,Gallo2011,
Grupe2012,Gallo2018}. In their low \xray\ 
states, they inevitably deviate significantly from the expected 
$\alpha_{\rm OX}\textrm{--}L_{2500~{\textup{\AA}}}$ ($L_{\rm 2~keV}\textrm{--}L_{2500~{\textup{\AA}}}$)
 relation. Their low \xray\ states can be explained 
by a \hbox{partial-covering} absorption scenario, 
where the geometrically inner
thick accretion disk and its associated outflow play the role of the 
absorber \citep[see][and references therein]{Luo2015,Liu2019,Ni2020}.
Analyses of their \hbox{low-state} spectra sometimes cannot 
 detect any absorption, perhaps because 
 they exhibit soft \hbox{$\approx 0.3\textrm{--}6$}~keV \xray\ 
 spectra, which are probably dominated by the \hbox{soft-excess}
  component or the transmitted fraction of \xray\ emission unaffected 
 by the partial covering absorber 
  \citep[see the case of SDSS~$\rm J075101.42+291419.1$ in]
  []{Liu2019}.

The fraction of extremely \xray\ variable AGNs 
among \hbox{super-Eddington} accreting AGNs was estimated to be 
\hbox{$\sim 15\textrm{--}24\%$} \citep{Liu2019}. 
In this study,
the three extremely \xray\ variable AGNs constitute a fraction of 
\hbox{$\approx 14\%$} (3/21) among the \hbox{super-Eddington} subsample, 
which is generally consistent with that reported in \cite{Liu2019}.
 Moreover, a number of our super-Eddington accreting AGNs, 
 mostly the SDSS quasars, have limited numbers (one or two) of 
\xray\ observations, and thus their \xray\ variability behavior is not
well constrained. It is thus possible that 
the actual number of extremely \xray\ 
variable AGNs among our sample is larger, resulting in a more 
 significant impact of \xray\ variability on the 
study of the $\alpha_{\rm OX}\textrm{--}L_{2500~{\textup{\AA}}}$ ($L_{\rm 2~keV}\textrm{--}L_{2500~{\textup{\AA}}}$) and $\Gamma\textrm{--}L_{\rm Bol}/
L_{\rm Edd}$ relations for \hbox{super-Eddington} accreting AGNs. 
We therefore emphasize the importance of using  
\hbox{high-state} \xray\ data to probe the intrinsic 
accretion disk-corona connections in AGNs, especially for objects
 with high accretion rates. Multiple \xray\ observations 
are required to collect the variability information for every sample 
object, in order to construct an unbiased sample for such studies.

\section{Summary and Future Prospects}
\label{sec:sum}
In this study, we constructed a sample of 47 AGNs with RM 
measurements, to systematically study the observational
 differences between the coronae and accretion disk-corona 
connections in super- and \hbox{sub-Eddington} accreting AGNs.
All our sample objects have sensitive \xray\ coverage 
from archival \xmm, \chandra, or \swift\ observations, and we have 
selected high-state data for objects with multiple observations to 
probe their intrinsic \xray\ emission. All the sample objects, 
except one, have simultaneous UV/optical data.
Our full sample was divided into the super-Eddington subsample with 
 $\dot{\rm \mathscr{M}}\ge 3$ and \hbox{sub-Eddington} subsample 
 with $\dot{\rm \mathscr{M}}<3$, 
 and we performed detailed statistical analysis on 
the $\alpha_{\rm OX}(L_{\rm 2~keV})
\textrm{--}L_{2500~{\textup{\AA}}}$ and 
$\Gamma\textrm{--}L_{\rm Bol}/L_{\rm Edd}(\dot{\rm \mathscr{M}})$ correlations for the two 
subsamples. Our main results are as follows:
\begin{enumerate}
\item
We found a strong correlation between $\alpha_{\rm OX}$ and 
$L_{2500~{\textup{\AA}}}$ for both the super- and \hbox{sub-Eddington}  
subsamples. The linear regression 
analysis on $\alpha_{\rm OX}$ versus  
${\rm log}L_{2500~{\textup{\AA}}}$ 
reveals a slope of $-0.106\pm 0.019$ for the \hbox{super-Eddington} 
subsample, which is slightly flatter than, but still consistent 
within $1\sigma$ with the slope of $-0.111\pm0.019$ for the 
\hbox{sub-Eddington} subsample. The best-fit intercepts are also consistent 
within $1\sigma$. A strong correlation between $L_{\rm 2~keV}$ 
and $L_{2500~{\textup{\AA}}}$ for both the 
super- and \hbox{sub-Eddington} subsamples was also found.
The best-fit ${\rm log}(L_{\rm 2~keV})
\textrm{--}{\rm log}(L_{2500~{\textup{\AA}}})$ relations for the two 
subsamples are also consistent considering the $1\sigma$ 
uncertainties.  
See Section~\ref{subsec:aox-l2500}.

\item
A statistically significant correlation was found between the hard 
(rest-frame $>2$~keV) \xray\ spectral photon index ($\Gamma$) and 
$L_{\rm Bol}/L_{\rm Edd}$ for the \hbox{super-Eddington}
 subsample and the full sample. 
 However, there is no significant correlation between $\Gamma$ 
 and $L_{\rm Bol}/L_{\rm Edd}$ for the \hbox{sub-Eddington} subsample. 
The slope of the best-fit $\Gamma\textrm{--}{\rm log}(L_{\rm Bol}/L_{\rm Edd})$
relation for the \hbox{super-Eddington} subsample is 
$0.33\pm 0.13$. See Section~\ref{subsec:ga-mdot}.

\item
We constructed IR-to-X-ray SEDs for ten
 \hbox{super-Eddington} accreting 
AGNs with $2500~{\textup{\AA}}$ luminosities  
 exceeding $10^{44.5}~\rm erg~s^{-1}$.
The SEDs of most objects are largely consistent with those of 
typical quasars, except for three objects that show weaker 
mid-IR emission. See Section~\ref{subsec:sed}.

\item
Super- and \hbox{sub-Eddington} accreting AGNs follow the same
$\alpha_{\rm OX}\textrm{--}L_{2500~{\textup{\AA}}}$ 
($L_{\rm 2~keV}\textrm{--}L_{2500~{\textup{\AA}}}$) relation, 
indicating that \hbox{super-Eddington} accreting AGNs are not significantly  
\xray\ weak compared to \hbox{sub-Eddington} accreting AGNs, as long 
as their intrinsic 
\xray\ emission is considered. These two groups   
likely share the same 
accretion power dissipation relation for the 
accretion disk-corona system. See Section~\ref{subsec:dis-aox}.

\item
We discuss the possibility that the $\alpha_{\rm OX}$ versus 
$L_{\rm Bol}/L_{\rm Edd}$ plus $M_{\rm BH}$ relation serves as the 
physical driver for the observed $\alpha_{\rm OX}\textrm{--}L_{2500~{\textup{\AA}}}$ relation. Significant dependences of 
$\alpha_{\rm OX}$ on both $L_{\rm Bol}/L_{\rm Edd}$ and $M_{\rm BH}$ 
are confirmed for our sample. 
A multi-variate linear regression revealed the relation: 
 $\alpha_{\rm OX}=(-0.13 \pm 0.01){\rm log}(L_{\rm Bol}/L_{\rm Edd})-(0.10 \pm 0.01){\rm log}M_{\rm BH}-(0.69\pm 0.09)$.
See Section~\ref{subsec:aox-lledd}.



\end{enumerate}
There is not a significant correlation between $\Gamma$ and $L_{\rm Bol}/L_{\rm Edd}$ for the sub-Eddington subsample, 
probably due to the small sample size. We thus cannot constrain 
the difference between the $\Gamma\textrm{--}L_{\rm Bol}/L_{\rm Edd}$ 
relations for the two subsamples.
 From our parent sample,
we excluded another six objects without \xray\ observations and
six objects with \hbox{low-S/N} observations. 
Five of them have new \chandra\ observations, which 
will provide constraints on their \xray\ properties in the near 
future. If targeted observations with \xmm\ or \chandra\ are 
obtained for the other objects, we will have a larger sample after adding 
 these 12 objects.
 
 We may also consider a larger 
 sample from some ongoing or upcoming AGN RM projects. For example,
  the ongoing SDSS-RM project is the first dedicated multi-object 
  RM program, executed with the 
  SDSS-\uppercase\expandafter{\romannumeral3}
   Baryon Oscillation Spectroscopic Survey (BOSS) spectrograph 
  \citep[e.g.,][]{Shen2015}.
The primary goal of this project is to obtain RM measurements for 
$\ga 100$ quasars, which cover a wider luminosity and redshift 
range compared to previous RM AGN samples 
\citep[e.g.,][]{Shen2016,Grier2017,Grier2019}.
This program is accompanied by a large \xmm\ program 
(\hbox{XMM-RM}) that
completes the \xray\ coverage of the same field \citep{Liuteng2020}. 
However, we note that this \xmm\ program has a limited number of 
observations with limited exposure times, and some quasars are not 
detected. These \xray\ observations cannot provide  
tight constraints on the \xray\ properties, and they might also 
be biased against \xray\ weak/undetected objects.
Targeted observations with \hbox{higher-quality} data are 
still required.
 
 There are some upcoming multi-object RM campaigns.
  The SDSS-\uppercase\expandafter{\romannumeral5} Black Hole Mapper 
 (BHM) will further extend the number of RM AGNs to an 
 "industrial scale" \citep{Kollmeier2017}. 
 It will perform RM campaigns in a number of fields 
 including three of the four Deep-Drilling Fields 
 (DDFs; \hbox{XMM-LSS}, CDF-S, and COSMOS) of the Vera C. Rubin 
 Observatory Legacy Survey of Space and Time (LSST), and the total 
 number of RM AGNs is expected to be $>1000$.
 The 4-metre multi-Object Spectroscopic Telescope (4MOST) 
TiDES-RM program will perform RM campaigns on the four 
LSST DDFs (\hbox{XMM-LSS}, CDF-S, COSMOS, and ELAIS-S1), and 
it will target around 700 AGNs \citep{Brandt2018,Swann2019}. 
The additional high-quality photometry from LSST will improve the 
 accuracy of measured BH masses and the number of 
 reverberation-lag detections. Moreover, 
 the four DDFs have good \xray\ 
 coverage from completed or ongoing \xmm\ and \chandra\ observations
 (e.g., \citealt{Chen2018}; Ni Q. et al. in preparation).
 Therefore, these programs will provide promising RM AGN samples with 
 high-quality multiwavelength data to study the accretion disk-corona 
 connections in AGNs.

Some issues related to the multiwavelength properties of 
\hbox{super-Eddington} accreting AGNs remain unclear.
These AGNs are expected to have different SEDs
 compared to \hbox{sub-Eddington} accreting AGNs, with the 
 primary differences arising in the EUV  
 band \citep[e.g.,][]{Castell2016,Kubota2018}. 
 The upcoming Chinese Space Station Optical survey 
 (\hbox{CSS-OS}; e.g., \citealt{Zhan2018}) will perform a deep NUV to 
optical imaging survey utilizing the Multi-Channel Imager.
It will provide rest-frame EUV photometric observations
 for a large sample of high-redshift 
AGNs, which are valuable for studying the difference between the SEDs 
of super- and \hbox{sub-Eddington} accreting AGNs.

  Our investigation has suggested that super-Eddington accreting 
  AGNs tend to show extreme \xray\ variability. It is thus important to obtain 
 the \hbox{high-state} or intrinsic \xray\ data for 
 \hbox{super-Eddington} accreting AGNs when studying the disk-corona connection in 
 these systems.
 The nature of extreme \xray\ variability in \hbox{super-Eddington} 
 accreting AGNs is still not well understood. 
With the ongoing eROSITA \citep{Merloni2012,Predehl2017} \xray\ 
survey of AGNs, we will obtain more variability information for the
\hbox{super-Eddington} accreting AGNs that have limited numbers of 
\xray\ observations now. We will likely discover more sources with 
extreme \xray\ variability. Nearby luminous AGNs with high-quality 
mulitwavelength data are optimal samples for examining the physical 
scenario for such extreme behavior \citep[e.g.,][]{Liu2019}.
Moreover, a systematic monitoring program of a uniform AGN 
sample selected from the RM campaigns discussed above, is required to
constrain better the occurrence rate of extreme \xray\ variability in 
\hbox{super-Eddington} accreting AGNs.

~\\
We thank the referee, Belinda Wilkes, for helpful suggestions and detailed comments.
We thank Pu~	Du, Chen~Hu, Jian-Min Wang,
Qiusheng~Gu, Yong~Shi, and Zhiyuan~Li for 
helpful discussions. 
H.L. and B.L. acknowledge financial support from
the National Natural Science Foundation of China
grants 11991053 and 11673010 and National Key R\&D Program of China 
grant 2016YFA0400702.
H.L. acknowledges financial support from the program of China 
Scholarships Council (No. 201906190104) for her visit to the 
Pennsylvania State University.
W.N.B. and J.D.T acknowledge support from NASA ADP grant 
80NSSC18K0878.
S.C.G. thanks the Natural Science and Engineering 
Research Council of Canada for support.

\bibliographystyle{aasjournal}
\bibliography{ms}

\newpage
\appendix
In this section, we give notes on individual objects 
showing strong variability in the \xray\ and/or 
UV/optical bands. 

\section{Mrk 335}
Mrk 335, classified as a super-Eddington accreting AGN, 
is a well-studied bright AGN that was discovered to fall 
into a historically low \xray\ state in 2007, with the flux dropping 
by a factor of $\sim30$
compared to the 2000 and 2006 \xmm\ fluxes \citep{Grupe2007a}. 
Since then, it has been continuously monitored by \swift, but 
it has never fully recovered to the previous bright state \cite[e.g.,]
[]{Grupe2008a,Grupe2012,Gallo2018}.
Compared to the 2000 \xmm\ observation,
the 2--10~keV \xray\ flux of the 2006 \xmm\ observation 
is slightly larger, and the exposure time is much longer \citep{Grupe2008a}. 
 We thus used the 2006 \xmm\ data of Mrk 335 for our study.
 
 \section{IRAS F$12397+3333$}
 IRAS F$12397+3333$, a super-Eddington accreting AGN,
  was observed by \xmm\ in two consecutive 
 revolutions during 2005 June 20 and 23, and the \xray\ flux varied
  mildly between these two segments. The observations reveal 
 that it was affected by ionized absorption mainly in the
 soft (\hbox{rest-frame} $<2$~keV) \xray\ emission \citep{Dou2016}. 
 Our simple power-law fitting of  
the rest-frame $>2~\rm keV$ spectrum yields a steep photon 
index of $\Gamma=2.15\pm 0.02$, consistent with that reported in 
\cite{Dou2016}. 
It has a \chandra\ observation in 
2000 (Observation ID: 3000). The rest-frame 2--7 keV \chandra\ 
spectrum is well modeled by a 
flatter power law with $\Gamma=1.30\pm 0.06$. The 2--10 keV  
flux of the \chandra\ observation is slightly smaller 
than those of the \xmm\ observations, and the 0.5--2 keV flux is about half of the \xmm\ 
fluxes. Therefore, the ionized absorption during the \chandra\ 
observation is likely stronger.  
We thus used the high-state \xmm\ observation in our study.

 The UV/optical spectrum of IRAS F$12397+3333$ appears to be
  affected by intrinsic reddening, based on the Balmer decrements of 
 $\rm H\alpha/H\beta=5.71$ and 6.14 for the broad and narrow 
 lines, respectively \citep[e.g.,][]{Du2014}. We found 
 an unusual flattening of the optical/UV spectral slope at shorter wavelengths for 
 the OM UV/optical fluxes, which also indicates intrinsic 
 reddening. In order to correct for the intrinsic reddening, we converted the 
Balmer decrement of the broad lines to $E_{\rm B-V}$ assuming 
an intrinsic $\rm H\alpha/H\beta=3.06$ \citep{Dong2008} and 
a Galactic extinction curve \citep{Cardelli1989}. The estimated 
$E_{\rm B-V}$ value is 0.59. We note that the extinction does not affect 
the hard X-ray emission of IRAS F$12397+3333$ (see Section~\ref{subsec:spec})
 
\section{Mrk 382} 
Mrk~382, a super-Eddington accreting AGN, 
was observed by \chandra\ for 4.9~ks
on 2010 December 6 (Observation ID:
 13008). The rest-frame 2--7 keV \chandra\ spectrum is well described  
 by a Galactic absorption modified power law with 
 $\Gamma=1.25 \pm 0.16$. The extrapolation of this power law to the
  0.5--7 keV energy range reveals a small excess below 2 keV. 
 In comparison, the rest-frame $> 2$ keV 
\xmm\ spectrum (observed on 2011 November 2) is much steeper ($
\Gamma=2.17\pm0.02$), with the \hbox{2--10 keV} flux slightly 
larger than that of the \chandra\ observation.
The $0.3\textrm{--}2/(1+z)$ keV \xmm\ spectrum can be well fitted  
with a single \hbox{power-law} model ($\Gamma_{\rm s}=2.66\pm 
0.01$) modified by Galactic absorption.
The 0.5--2 keV flux was measured to be $4.89\times 
10^{-12}\ \rm erg\ cm^{-2}\ s^{-1}$, larger than the \chandra\ 0.5--2 keV flux 
($6.11 \times 10^{-13} \rm erg\ cm^{-2}\ s^{-1}$) by a factor of about eight.
Moreover, we have analyzed two archival \swift\ observations 
(2009 February 6 and 2011 September 12, respectively), which reveal 
fluxes between those in the \chandra\ and \xmm\ observations.
We thus considered the 2011 \xmm\ observation as the high-state observation.

\section{Mrk 493}
Mrk 493 is a super-Eddington accreting AGN.
\cite{Bonson2018} has reported that 
the flux of Mrk~493 observed in the 2015 \xmm\ observation 
is about half of that of the 2003 \xmm\ observation. 
Mrk 493 also has one \chandra\ observation (2010 February 7; \citealt{Dong2012}), 
during which the flux and spectral photon index are consistent with those of the 
2003 \xmm\ observation. 
We thus used the 2003 \xmm\ observation in our analysis.


\section{PG $1211+143$}
PG $1211+143$, classified as a super-Eddington accreting AGN, 
is also well known for its extreme \xray\ variability 
and spectral complexity, and it is an archetypical case of an AGN 
exhibiting an ultra-fast outflow 
\citep[e.g.,][]
{Pounds2003,Bachev2009,deMarco2011,Pounds2016,Lobban2016,Reeves2018}.
We analyzed additional \xmm\ and \swift\ archival observations that 
have not been reported in the literature, and compared the derived \xray\ fluxes 
with those reported in the aforementioned studies. 
We finally selected the highest-state observation 
(Observation ID: 0745110601) of the continuous \xmm\ 
observations in 2014 \citep[see e.g.,][]{Lobban2016}.

\section{PG $0844+349$}
PG $0844+349$ is a \hbox{super-Eddington} accreting AGN showing 
extreme \xray\ variability by a factor of larger than 10, and 
its long-term \xray\ light curve was presented in 
\cite{Gallo2011}.
The observation used in this study is its 2001 high-state \xmm\ 
observation.

\section{Mrk 1310}
Mrk 1310 is a \hbox{sub-Eddington} accreting AGN that has been  
observed multiple times by \xmm\ and \swift\ from 2006 to 2019. 
Its \xray\ flux showed 
extreme variability with a maximum amplitude of $\sim$ 30. 
Meanwhile, coordinated variability in UV/optical and IR fluxes 
was also observed, with smaller variability amplitudes toward longer 
wavelengths. Its optical spectral type has also changed, which makes
 it a changing-look AGN. The detailed analysis on this source will be
 reported in Luo B. et al. (in preparation).
In this study, we used its 2016 \swift\ observation when it exhibited 
the brightest multiwavelengh fluxes.

\section{NGC 2617}
NGC 2617 is a sub-Eddington accreting AGN that underwent a dramatic 
\xray\ outburst from 2013 April to May, during which its \xray\ flux 
increased by an order of magnitude, followed by an increase of the UV/
optical flux by almost an order of magnitude \citep[e.g.,][]
{Shappee2014,Giustini2017}. 
Its optical spectral type switched from Seyfert 1.8 to Seyfert 1.0 
due to the appearance of broad optical emission lines. 
It was observed by \xmm\ on 2013 May 24 
when its \xray\ flux was at the peak level
\cite[see Figure~4 of][]{Shappee2014}. 
We thus used this observation in our study.

\section{Arp~151} 
Arp~151 is a sub-Eddington accreting AGN that showed   
normal \xray\ emission in its 2009 February 15 \swift\ 
observation. The 0.3--10 keV spectrum can be well fitted by a single 
power-law model modified by Galactic absorption with $\Gamma=1.62\pm 0.09$, 
and the 0.5--2 keV flux was measured to be $4.80\times10^{-12}~\rm erg\ cm^{-2}\ s^{-1}$.
It was observed by \chandra\ on 2011 December 4 
(Observation ID: 12871), when it 
exhibited a flatter 0.5--7~keV spectrum with $\Gamma=1.25\pm 0.02$. 
The 0.5--2 keV flux ($1.47\times10^{-12}~\rm erg\ cm^{-2}\ s^{-1}$) 
decreased by a factor of $\approx 3$, compared to the \swift\ flux. 
We thus used the \swift\ observation in our study.

\clearpage

\startlongtable
\begin{deluxetable*}{lccccccccc}
\tablewidth{0pt}
\tablecaption{BH Masses, Accretion Rates, and Optical Properties
}
\tablehead{
\colhead{Object}&
\colhead{$z$}&
\colhead{$E_{\rm B-V}$}& 
\colhead{$N_{\rm H,gal}$}&
\colhead{log $L_{\rm 5100~{\textup{\AA}}}$} &
\colhead{FWHM($\rm H\beta$)} &
\colhead{log($M_{\rm BH}/M_\odot$)} &
\colhead{log~$\rm\dot{\mathscr{M}}$}  &
\colhead{log~$L_{\rm Bol}$}  &
\colhead{log~$L_{\rm Bol}/L_{\rm Edd}$}  \\
\colhead{   } &
\colhead{ } &
\colhead{ } &
\colhead{($\rm 10^{20}\ cm^{-2}$) } &
\colhead{($\rm erg\ s^{-1}$)}  &
\colhead{($\rm km\ s^{-1}$) } &
\colhead{ } &
\colhead{ } &
\colhead{ ($\rm erg\ s^{-1}$)} &
\colhead{    } \\
\colhead{ (1) } &
\colhead{ (2) } &
\colhead{ (3) } &
\colhead{(4)} &
\colhead{(5)}  &
\colhead{(6) } &
\colhead{(7) } &
\colhead{(8) } &
\colhead{ (9)  } &
\colhead{ (10)  }       
}
\startdata
\multicolumn{10}{c}{\hbox{Super-Eddington} Subsample ($\rm\dot{\mathscr{M}}\ge 3$)} \\
 \midrule
Mrk $335$&$0.026$&$0.035$&$3.48$&$43.76$&$1707$&$6.93^{+0.10}_{-0.11}$&$1.20^{+0.22}_{-0.20}$&$44.88$&$-0.23^{+0.11}_{-0.10}$\\
Mrk $1044$&$0.016$&$0.033$&$3.10$&$43.10$&$1178$&$6.45^{+0.12}_{-0.13}$&$1.50^{+0.26}_{-0.24}$&$44.48$&$-0.15^{+0.13}_{-0.12}$\\
IRAS $04416+1215$&$0.089$&$0.436$&$12.43$&$44.47$&$1522$&$6.78^{+0.31}_{-0.06}$&$2.72^{+0.11}_{-0.62}$&$45.52$&$0.57^{+0.08}_{-0.32}$\\
SDSS J$074352.02+271239.5$&$0.252$&$0.045$&$4.69$&$45.37$&$3156$&$7.93^{+0.05}_{-0.04}$&$1.59^{+0.09}_{-0.10}$&$46.32$&$0.21^{+0.10}_{-0.10}$\\
Mrk $382$&$0.034$&$0.048$&$4.81$&$43.12$&$1462$&$6.50^{+0.19}_{-0.29}$&$1.25^{+0.58}_{-0.38}$&$44.38$&$-0.29^{+0.29}_{-0.19}$\\
SDSS J$080101.41+184840.7$&$0.140$&$0.032$&$2.53$&$44.27$&$1930$&$6.78^{+0.34}_{-0.17}$&$2.22^{+0.34}_{-0.67}$&$45.31$&$0.35^{+0.17}_{-0.34}$\\
SDSS J$081441.91+212918.5$&$0.163$&$0.039$&$3.50$&$43.96$&$1729$&$7.18^{+0.20}_{-0.20}$&$1.36^{+0.40}_{-0.40}$&$45.30$&$-0.06^{+0.20}_{-0.20}$\\
PG $0844+349$&$0.064$&$0.036$&$2.79$&$44.22$&$2694$&$7.66^{+0.15}_{-0.23}$&$0.60^{+0.47}_{-0.31}$&$45.37$&$-0.47^{+0.23}_{-0.15}$\\
SDSS J$085946.35+274534.8$&$0.244$&$0.032$&$2.51$&$44.41$&$1718$&$7.30^{+0.19}_{-0.61}$&$1.29^{+1.22}_{-0.38}$&$45.28$&$-0.20^{+0.61}_{-0.20}$\\
Mrk $110$&$0.035$&$0.013$&$1.32$&$43.66$&$1633$&$7.10^{+0.13}_{-0.14}$&$1.00^{+0.28}_{-0.26}$&$45.00$&$-0.28^{+0.14}_{-0.13}$\\
SDSS J$093922.89+370943.9$&$0.186$&$0.014$&$1.19$&$44.07$&$1209$&$6.53^{+0.07}_{-0.33}$&$2.45^{+0.66}_{-0.14}$&$45.16$&$0.45^{+0.33}_{-0.09}$\\
PG $0953+414$&$0.234$&$0.012$&$1.25$&$45.19$&$3070$&$8.44^{+0.06}_{-0.07}$&$0.53^{+0.14}_{-0.12}$&$46.34$&$-0.28^{+0.07}_{-0.06}$\\
SDSS J$100402.61+285535.3$&$0.329$&$0.022$&$1.77$&$45.52$&$2088$&$7.44^{+0.37}_{-0.06}$&$2.56^{+0.12}_{-0.74}$&$46.34$&$0.73^{+0.06}_{-0.37}$\\
Mrk $142$&$0.045$&$0.016$&$1.28$&$43.59$&$1462$&$6.47^{+0.38}_{-0.38}$&$1.63^{+0.76}_{-0.76}$&$44.49$&$-0.15^{+0.38}_{-0.38}$\\
UGC $6728$&$0.007$&$0.100$&$4.37$&$41.86$&$1641$&$5.87^{+0.19}_{-0.40}$&$0.64^{+0.81}_{-0.37}$&$43.05$&$-0.99^{+0.40}_{-0.19}$\\
PG $1211+143$&$0.081$&$0.033$&$2.92$&$44.73$&$2012$&$7.87^{+0.11}_{-0.26}$&$0.55^{+0.52}_{-0.21}$&$45.60$&$-0.44^{+0.26}_{-0.11}$\\
IRASF $12397+3333$&$0.043$&$0.019$&$1.42$&$44.23$&$1802$&$6.79^{+0.27}_{-0.45}$&$1.80^{+0.91}_{-0.53}$&$44.97$&$0.00^{+0.45}_{-0.27}$\\
NGC $4748$&$0.015$&$0.051$&$3.50$&$42.56$&$1947$&$6.61^{+0.11}_{-0.23}$&$0.57^{+0.45}_{-0.23}$&$44.03$&$-0.76^{+0.23}_{-0.11}$\\
Mrk $493$&$0.031$&$0.025$&$2.09$&$43.11$&$778$&$6.14^{+0.04}_{-0.11}$&$1.88^{+0.22}_{-0.09}$&$44.35$&$0.03^{+0.11}_{-0.05}$\\
KA $1858+4850$&$0.079$&$0.054$&$4.28$&$43.43$&$1820$&$6.94^{+0.07}_{-0.09}$&$0.90^{+0.19}_{-0.14}$&$44.67$&$-0.45^{+0.13}_{-0.11}$\\
PG $2130+099$&$0.062$&$0.043$&$3.80$&$44.20$&$2450$&$7.05^{+0.08}_{-0.10}$&$1.98^{+0.19}_{-0.16}$&$45.44$&$0.21^{+0.10}_{-0.08}$\\
\midrule
\multicolumn{10}{c}{\hbox{Sub-Eddington} Subsample ($\rm\dot{\mathscr{M}}< 3$)} \\
 \midrule
PG $0026+129$&$0.142$&$0.072$&$4.56$&$44.97$&$2543$&$8.15^{+0.09}_{-0.13}$&$0.18^{+0.26}_{-0.17}$&$45.74$&$-0.59^{+0.13}_{-0.09}$\\
PG $0052+251$&$0.154$&$0.046$&$4.40$&$44.81$&$5007$&$8.64^{+0.11}_{-0.14}$&$-0.53^{+0.27}_{-0.21}$&$45.95$&$-0.87^{+0.14}_{-0.11}$\\
Fairall $9$&$0.047$&$0.025$&$3.22$&$43.98$&$5998$&$8.09^{+0.07}_{-0.12}$&$-0.40^{+0.25}_{-0.15}$&$45.34$&$-0.92^{+0.12}_{-0.07}$\\
Ark $120$&$0.033$&$0.128$&$9.70$&$43.87$&$6077$&$8.47^{+0.07}_{-0.08}$&$-1.05^{+0.16}_{-0.14}$&$45.36$&$-1.29^{+0.08}_{-0.07}$\\
MCG $+08-11-011$&$0.020$&$0.214$&$18.48$&$43.33$&$4138$&$7.72^{+0.04}_{-0.05}$&$-0.83^{+0.10}_{-0.09}$&$44.56$&$-1.34^{+0.05}_{-0.04}$\\
Mrk $374$&$0.043$&$0.052$&$5.16$&$43.77$&$4980$&$7.86^{+0.15}_{-0.12}$&$-0.80^{+0.24}_{-0.29}$&$44.67$&$-1.36^{+0.12}_{-0.15}$\\
Mrk $79$&$0.022$&$0.071$&$5.06$&$43.68$&$4793$&$7.84^{+0.12}_{-0.16}$&$-1.31^{+0.32}_{-0.24}$&$44.46$&$-1.56^{+0.16}_{-0.12}$\\
PG $0804+761$&$0.064$&$0.035$&$3.29$&$44.91$&$3052$&$8.43^{+0.05}_{-0.06}$&$-0.27^{+0.12}_{-0.11}$&$45.78$&$-0.82^{+0.06}_{-0.05}$\\
NGC $2617$&$0.014$&$0.034$&$3.52$&$42.67$&$8026$&$7.74^{+0.11}_{-0.17}$&$-1.39^{+0.35}_{-0.21}$&$44.30$&$-1.62^{+0.17}_{-0.11}$\\
SBS $1116+583A$&$0.028$&$0.011$&$0.75$&$42.14$&$3668$&$6.78^{+0.11}_{-0.12}$&$-0.08^{+0.23}_{-0.22}$&$43.79$&$-1.17^{+0.12}_{-0.11}$\\
Arp $151$&$0.021$&$0.014$&$1.03$&$42.55$&$3098$&$6.87^{+0.05}_{-0.08}$&$-0.13^{+0.17}_{-0.11}$&$43.90$&$-1.15^{+0.09}_{-0.06}$\\
MCG $+06-26-012$&$0.033$&$0.019$&$1.93$&$42.67$&$1334$&$6.92^{+0.14}_{-0.12}$&$-0.14^{+0.23}_{-0.27}$&$43.96$&$-1.14^{+0.15}_{-0.16}$\\
Mrk $1310$&$0.020$&$0.031$&$2.50$&$42.29$&$2409$&$6.62^{+0.07}_{-0.08}$&$0.38^{+0.16}_{-0.13}$&$43.91$&$-0.88^{+0.09}_{-0.08}$\\
Mrk $766$&$0.013$&$0.020$&$1.86$&$42.57$&$1609$&$6.49^{+0.10}_{-0.10}$&$-0.10^{+0.20}_{-0.21}$&$43.67$&$-1.00^{+0.10}_{-0.10}$\\
Ark $374$&$0.063$&$0.027$&$2.82$&$43.70$&$3827$&$8.03^{+0.10}_{-0.08}$&$-0.80^{+0.15}_{-0.21}$&$44.96$&$-1.25^{+0.08}_{-0.10}$\\
NGC $4593$&$0.009$&$0.025$&$1.87$&$42.62$&$5141$&$7.26^{+0.09}_{-0.09}$&$-1.23^{+0.18}_{-0.18}$&$43.66$&$-1.78^{+0.09}_{-0.09}$\\
PG $1307+085$&$0.155$&$0.034$&$2.26$&$44.85$&$5058$&$8.72^{+0.13}_{-0.26}$&$-0.90^{+0.51}_{-0.26}$&$45.70$&$-1.20^{+0.26}_{-0.13}$\\
Mrk $279$&$0.030$&$0.016$&$1.67$&$43.71$&$5354$&$7.97^{+0.09}_{-0.12}$&$-0.89^{+0.23}_{-0.18}$&$44.89$&$-1.26^{+0.12}_{-0.09}$\\
NGC $5548$&$0.017$&$0.020$&$1.65$&$43.29$&$7241$&$8.10^{+0.16}_{-0.16}$&$-1.93^{+0.32}_{-0.32}$&$44.36$&$-1.92^{+0.16}_{-0.16}$\\
PG $1426+015$&$0.087$&$0.032$&$2.58$&$44.63$&$7112$&$8.97^{+0.12}_{-0.22}$&$-1.27^{+0.43}_{-0.24}$&$45.85$&$-1.30^{+0.22}_{-0.12}$\\
Mrk $817$&$0.031$&$0.007$&$1.16$&$43.74$&$5347$&$7.99^{+0.14}_{-0.14}$&$-0.65^{+0.28}_{-0.28}$&$44.98$&$-1.18^{+0.14}_{-0.14}$\\
Mrk $290$&$0.030$&$0.014$&$1.55$&$43.17$&$4542$&$7.55^{+0.07}_{-0.07}$&$-0.48^{+0.15}_{-0.14}$&$44.52$&$-1.20^{+0.07}_{-0.07}$\\
Mrk $876$&$0.139$&$0.027$&$2.42$&$44.77$&$9073$&$8.81^{+0.14}_{-0.21}$&$-0.66^{+0.41}_{-0.28}$&$45.99$&$-1.00^{+0.21}_{-0.14}$\\
NGC $6814$&$0.005$&$0.184$&$9.08$&$42.12$&$3323$&$7.16^{+0.05}_{-0.06}$&$-1.70^{+0.13}_{-0.11}$&$43.18$&$-2.16^{+0.06}_{-0.05}$\\
Mrk $509$&$0.034$&$0.057$&$4.17$&$44.19$&$3014$&$8.15^{+0.03}_{-0.03}$&$-0.26^{+0.06}_{-0.06}$&$45.46$&$-0.87^{+0.03}_{-0.03}$\\
NGC $7469$&$0.016$&$0.070$&$4.17$&$43.51$&$4369$&$7.60^{+0.12}_{-0.06}$&$-0.49^{+0.11}_{-0.24}$&$44.58$&$-1.20^{+0.06}_{-0.12}$\\\enddata
\label{tbl1}
\tablecomments{Columns (1), (2): object name and redshift.
Column (3): Galactic extinction from \cite{Schlegel1998}.
Column (4): Galactic neutral hydrogen column density from \cite{Dickey1990}.
Columns (5)--(7): $5100~\textup{\AA}$ luminosity, H$\beta$ FWHM, and BH mass, adopted from \cite{Du2015,Du2016,Du2019}.
Column (8): normalized accretion rate calculated using $\dot{\rm \mathscr{M}}= 4.82 (\ell_{44}/{\cos i})^{3/2} m_7^{-2}$ ($\cos~i=0.75$ adopted).
Column (9): bolometric luminosity, measured through integrating the 
IR-to-X-ray SED (see Section~\ref{subsec:bol} for details). The uncertainties on the bolometric luminosities range from 0.001--0.09 dex, with a median value of 0.01 dex. 
Column (10): Eddington ratio.
}
\end{deluxetable*}

\begin{longrotatetable}
\begin{deluxetable*}{lcccccccccccr}
\tablewidth{0pt}
\tablecaption{X-ray Observation Log and Hard X-ray Spectral Fitting Results}
\tablehead{
\colhead{Object}&
\colhead{Observatory}& 
\colhead{Observation} &
\colhead{Exposure} &
\colhead{Source} &
\colhead{$\Gamma$}  &
\colhead{${\rm log}\ f_{\rm 2~keV}$}  &
\colhead{log $L_{\rm 2\textrm{--}10~keV}$}&
\colhead{W-stat/dof}  &
\colhead{$N_{\rm UV}$}  &
\colhead{log $L_{\rm 2500~{\textup{\AA}}}$}&
\colhead{$\alpha_{\rm OX}$} &
\colhead{$\Delta\alpha_{\rm OX}$} \\
\colhead{   } &
\colhead{ }&
\colhead{ID}  &
\colhead{Time (ks)} &
\colhead{Counts} & 
\colhead{ } &
\colhead{ ($\rm erg\ cm ^{-2}\ s^{-1}\ Hz^{-1}$)} &
\colhead{($\rm erg\ s^{-1}$) }  &
\colhead{ }  &
\colhead{ }  & 
\colhead{($\rm erg\ s^{-1}\ Hz^{-1}$) }  &
\colhead{}  &
\colhead{} \\
\colhead{ (1) }&
\colhead{ (2) }&
\colhead{ (3) }&
\colhead{ (4) }&
\colhead{ (5) }&
\colhead{ (6) }&
\colhead{ (7) }&
\colhead{ (8) }&
\colhead{ (9) }&
\colhead{ (10) }&
\colhead{ (11) }&
\colhead{ (12) }&
\colhead{ (13) } 
}
\startdata
\multicolumn{13}{c}{\hbox{Super-Eddington} Subsample ($\rm\dot{\mathscr{M}}\ge 3$)} \\
 \midrule
Mrk $335$\tablenotemark{$\rm b$}&X&0306870101&89.7&$179984$&$2.10^{+0.01}_{-0.01}$&$-28.60$&$43.44$&$1526.7/1216$&$-1$&$29.05$&$-1.33$&$0.01$\\
Mrk $1044$&X&0824080401&97.0&$164068$&$2.27^{+0.01}_{-0.01}$&$-28.63$&$42.96$&$1395.5/1209$&1&$28.62$&$-1.33$&$-0.04$\\
IRAS $04416+1215$&S&00046111001&4.4&$125$&$2.46^{+0.27}_{-0.26}$&$-29.37$&$43.63$&$106.5/99$&1&$29.86$&$-1.52$&$-0.07$\\
SDSS J$074352.02+271239.5$&S&00046112004&3.4&$73$&$2.06^{+0.31}_{-0.30}$&$-29.63$&$44.44$&$66.4/66$&1&$30.64$&$-1.56$&$-0.00$\\
Mrk $382$\tablenotemark{$\rm a$}&X&0670040101&25.8&$12318$&$2.17^{+0.02}_{-0.02}$&$-29.25$&$43.00$&$1292.8/1379$&2&$28.50$&$-1.28$&$-0.02$\\
SDSS J$080101.41+184840.7$&X&0761510201&19.8&$2407$&$2.33^{+0.06}_{-0.06}$&$-29.86$&$43.59$&$681.0/864$&3&$29.53$&$-1.43$&$-0.02$\\
SDSS J$081441.91+212918.5$&X&0761510301&30.9&$3778$&$2.17^{+0.04}_{-0.04}$&$-29.89$&$43.75$&$854.5/1032$&2&$29.49$&$-1.37$&$0.03$\\
PG $0844+349$\tablenotemark{$\rm a,\ b$}&X&0103660201&5.9&$3507$&$2.22^{+0.05}_{-0.05}$&$-29.11$&$43.68$&$691.9/793$&$-1$&$29.62$&$-1.44$&$-0.02$\\
SDSS J$085946.35+274534.8$&C&5676&7.4&$138$&$1.82^{+0.23}_{-0.22}$&$-30.36$&$43.77$&$71.5/100$&$0$&$29.60$&$-1.46$&$-0.04$\\
Mrk $110$&X&0201130501&32.9&$64426$&$1.79^{+0.01}_{-0.01}$&$-28.50$&$43.92$&$1339.0/1219$&3&$29.14$&$-1.22$&$0.13$\\
SDSS J$093922.89+370943.9$&X&0411980301&4.2&$190$&$2.42^{+0.29}_{-0.25}$&$-30.29$&$43.39$&$144.0/175$&1&$29.35$&$-1.43$&$-0.04$\\
PG $0953+414$&X&0111290201&10.6&$5618$&$2.02^{+0.03}_{-0.03}$&$-29.30$&$44.72$&$1015.2/1174$&3&$30.62$&$-1.45$&$0.10$\\
SDSS J$100402.61+285535.3$&X&0804560201&40.5&$3915$&$2.31^{+0.05}_{-0.05}$&$-29.94$&$44.29$&$803.2/809$&2&$30.63$&$-1.59$&$-0.03$\\
Mrk $142$&S&00036539001&12.5&$212$&$2.12^{+0.25}_{-0.24}$&$-29.59$&$42.93$&$122.0/140$&3&$28.72$&$-1.40$&$-0.10$\\
UGC $6728$&S&00088256001&6.9&$928$&$1.64^{+0.09}_{-0.09}$&$-28.88$&$42.12$&$342.8/384$&1&$27.26$&$-1.21$&$-0.12$\\
PG $1211+143$\tablenotemark{$\rm b$}&X&0745110501&50.9&$25632$&$2.07^{+0.02}_{-0.02}$&$-29.21$&$43.84$&$1136.7/1186$&3&$29.87$&$-1.49$&$-0.04$\\
IRASF $12397+3333$\tablenotemark{$\rm a$}&X&0202180201&55.0&$32163$&$2.15^{+0.02}_{-0.02}$&$-29.13$&$43.34$&$1156.3/1177$&3&$29.26$&$-1.44$&$-0.07$\\
NGC $4748$\tablenotemark{$\rm a$}&X&0723100401&26.3&$14909$&$2.07^{+0.02}_{-0.02}$&$-28.95$&$42.60$&$960.6/1052$&2&$28.20$&$-1.33$&$-0.11$\\
Mrk $493$\tablenotemark{$\rm a$}&X&0112600801&12.2&$6120$&$2.18^{+0.03}_{-0.03}$&$-29.23$&$42.95$&$1191.9/1208$&3&$28.45$&$-1.28$&$-0.02$\\
KA $1858+4850$&S&00041251016&2.3&$50$&$1.92^{+0.48}_{-0.47}$&$-29.62$&$43.47$&$36.9/42$&1&$28.86$&$-1.27$&$0.04$\\
PG $2130+099$&X&0744370201&21.7&$1843$&$2.04^{+0.07}_{-0.07}$&$-29.17$&$43.66$&$598.0/664$&2&$29.73$&$-1.51$&$-0.08$\\
\midrule
\multicolumn{13}{c}{\hbox{Sub-Eddington} Subsample ($\rm\dot{\mathscr{M}}< 3$)} \\
 \midrule
PG $0026+129$&X&0783270201&5.7&$2779$&$1.75^{+0.05}_{-0.05}$&$-29.32$&$44.35$&$728.6/794$&2&$29.99$&$-1.39$&$0.08$\\
PG $0052+251$\tablenotemark{$\rm a$}&X&0301450401&13.7&$7193$&$1.75^{+0.03}_{-0.03}$&$-29.10$&$44.64$&$1057.8/1120$&1&$30.18$&$-1.35$&$0.15$\\
Fairall $9$&X&0741330101&5.2&$8484$&$1.91^{+0.03}_{-0.03}$&$-28.55$&$44.08$&$913.1/1026$&2&$29.53$&$-1.29$&$0.11$\\
Ark $120$&X&0721600201&86.2&$351705$&$1.96^{+0.00}_{-0.00}$&$-28.31$&$43.98$&$1752.4/1222$&3&$29.60$&$-1.35$&$0.06$\\
MCG $+08-11-011$&X&0201930201&26.4&$109848$&$1.64^{+0.01}_{-0.01}$&$-28.36$&$43.64$&$1317.0/1212$&2&$28.75$&$-1.20$&$0.10$\\
Mrk $374$&S&00037356003&12.7&$647$&$1.83^{+0.11}_{-0.11}$&$-29.29$&$43.28$&$266.8/321$&1&$28.95$&$-1.39$&$-0.06$\\
Mrk $79$&X&0400070201&13.9&$21739$&$1.86^{+0.02}_{-0.02}$&$-28.56$&$43.42$&$1190.5/1164$&2&$28.59$&$-1.19$&$0.09$\\
PG $0804+761$&X&0605110101&16.2&$11784$&$2.08^{+0.03}_{-0.03}$&$-28.87$&$43.97$&$987.4/1064$&2&$30.06$&$-1.52$&$-0.04$\\
NGC $2617$&X&0701981901&14.8&$70024$&$1.76^{+0.01}_{-0.01}$&$-28.29$&$43.35$&$1286.9/1207$&1&$28.40$&$-1.16$&$0.09$\\
SBS $1116+583\rm A$\tablenotemark{$\rm a$}&X&0821871801&18.7&$3104$&$1.72^{+0.05}_{-0.05}$&$-29.71$&$42.53$&$892.2/1045$&2&$28.00$&$-1.33$&$-0.13$\\
Arp $151$&S&00037369002&9.2&$972$&$1.62^{+0.09}_{-0.09}$&$-29.02$&$43.01$&$347.1/400$&3&$28.09$&$-1.19$&$0.02$\\
MCG $+06-26-012$&S&00040821003&4.8&$98$&$1.95^{+0.31}_{-0.30}$&$-29.57$&$42.73$&$80.5/84$&1&$28.14$&$-1.27$&$-0.06$\\
Mrk $1310$&S&00081108002&2.9&$256$&$1.86^{+0.18}_{-0.18}$&$-29.03$&$42.85$&$156.1/185$&2&$28.09$&$-1.22$&$-0.01$\\
Mrk $766$&X&0096020101&25.5&$41016$&$1.94^{+0.01}_{-0.01}$&$-28.72$&$42.77$&$1175.7/1190$&3&$27.60$&$-1.05$&$0.09$\\
Ark $374$&X&0301450201&17.2&$4733$&$1.94^{+0.04}_{-0.04}$&$-29.38$&$43.50$&$1063.8/1174$&1&$29.18$&$-1.38$&$-0.02$\\
NGC $4593$&X&0784740101&98.5&$135063$&$1.67^{+0.01}_{-0.01}$&$-28.66$&$42.61$&$1254.4/1203$&1&$27.86$&$-1.25$&$-0.07$\\
PG $1307+085$&X&0110950401&9.9&$1583$&$1.50^{+0.06}_{-0.06}$&$-29.78$&$44.06$&$631.4/811$&2&$30.04$&$-1.56$&$-0.08$\\
Mrk $279$&X&0302480401&40.4&$72230$&$1.82^{+0.01}_{-0.01}$&$-28.52$&$43.76$&$1299.4/1220$&3&$29.04$&$-1.24$&$0.10$\\
NGC $5548$&X&0089960301&58.2&$218513$&$1.63^{+0.01}_{-0.01}$&$-28.42$&$43.43$&$1300.0/1210$&$-1$&$28.52$&$-1.20$&$0.07$\\
PG $1426+015$\tablenotemark{$\rm a$}&X&0102040501&5.3&$5945$&$1.96^{+0.03}_{-0.03}$&$-28.95$&$44.20$&$927.1/1002$&1&$30.12$&$-1.47$&$0.02$\\
Mrk $817$\tablenotemark{$\rm a$}&X&0601781401&5.0&$8127$&$2.05^{+0.03}_{-0.03}$&$-28.75$&$43.48$&$1172.4/1279$&2&$29.23$&$-1.39$&$-0.02$\\
Mrk $290$&X&0400360201&13.9&$14545$&$1.52^{+0.02}_{-0.02}$&$-29.07$&$43.30$&$1435.8/1517$&1&$28.75$&$-1.35$&$-0.05$\\
Mrk $876$&X&0102040601&3.3&$1927$&$1.62^{+0.07}_{-0.07}$&$-29.34$&$44.36$&$765.6/814$&2&$30.31$&$-1.53$&$-0.01$\\
NGC $6814$&X&0550451801&9.2&$19109$&$1.64^{+0.02}_{-0.02}$&$-28.66$&$42.14$&$1086.0/1146$&$-1$&$27.41$&$-1.26$&$-0.14$\\
Mrk $509$&X&0601390201&40.0&$176765$&$1.80^{+0.01}_{-0.01}$&$-28.32$&$44.08$&$1512.3/1222$&3&$29.70$&$-1.38$&$0.05$\\
NGC $7469$&X&0207090101&59.2&$173305$&$1.89^{+0.01}_{-0.01}$&$-28.47$&$43.24$&$1500.6/1209$&3&$28.82$&$-1.35$&$-0.04$
\enddata
\tablecomments{
Column (1): object name. 
Column (2): \xray\ Observatory. X: \xmm; C: \chandra; S: \swift.
Column (3): observation ID.
Column (4): cleaned exposure time after filtering for high-background flares.
Column (5): number of rest-frame $> 2~\rm keV$
background-subtracted source counts used for spectral fitting.
Column (6): photon index obtained from spectral fitting of the 
rest-frame $>2$~keV spectrum using a power-law model modified by Galactic absorption.
{Columns (7), (8): logarithms of the flux density at rest-frame 2~keV and rest-frame 2--10 keV luminosity.}
Column (9): ratio between the W-statistic value and the number of degrees of freedom.
{Column (10): Number of available UV filters. ``$-2$'' denotes that only grism spectral data are available. ``$-1$'' denotes that 
UV-filter data are not available and U-filter data were used for calculating the flux density at $2500~\textup{\AA}$. ``0'' denotes that 
there are no simultaneous UV/optical data for the only \chandra\ object, SDSS J$085946.35+274534.8$, for which the flux density at $2500~\textup{\AA}$ was interpolated from its {\it GALEX} NUV and SDSS $u$-band flux densities.}
Column (11): logarithm of the rest-frame $2500~{\textup{\AA}}$ monochromatic luminosity.
Columns (12), (13): observed $\alpha_{\rm OX}$ value, and the difference between the observed $\alpha_{\rm OX}$ and the expected value derived from the \cite{Steffen2006} $\alpha_{\rm OX}\textrm{--}L_{\rm 2500~{\textup{\AA}}}$ relation. 
{\tablenotetext{$\rm a$}{X-ray sources affected by pile-up.} 
\tablenotetext{$\rm b$}{Objects that have shown extreme X-ray variability by factors of larger than 10.}} 
%
}
\label{tbl2}
\end{deluxetable*}
\end{longrotatetable}

\begin{longrotatetable}
\begin{deluxetable*}{lccccccccc}
\tablecaption{Soft X-ray Spectral Fitting Results}
\tablehead{
\colhead{Object}&
\colhead{Model}&
\colhead{$\Gamma_{\rm s}$}&
\multicolumn{3}{c}{{\sc comptt}}  &
\multicolumn{3}{c}{{\sc zxipcf}}  &
\colhead{log $L_{0.3\textrm{--}2~\rm keV}$} \\
\cmidrule(l){4-6} \cmidrule(l){7-9}
\colhead{ } &
\colhead{ } &
\colhead{ } &
\colhead{$T_{\rm seed}$ (eV)}&
\colhead{$kT$ (keV)}& 
\colhead{$\tau$}&
\colhead{$N_{\rm H}$ $\rm (10^{22}\ cm^{-2})$} &
\colhead{log $\xi$ } &
\colhead{$f_{\rm cov}$} &
\colhead{($\rm erg\ s^{-1}$) } \\
\colhead{ (1) } &
\colhead{ (2) } &
\colhead{ (3) } &
\colhead{(4)} &
\colhead{(5)} &
\colhead{(6)} &
\colhead{(7)} &
\colhead{(8)}&
\colhead{(9)} & 
\colhead{(10)}   
}
\startdata
\multicolumn{10}{c}{\hbox{Super-Eddington} Subsample ($\rm\dot{\mathscr{M}}\ge 3$)} \\
 \midrule
Mrk $335$&B&2.10&$67\pm 1$&$0.230^{+0.004}_{-0.004}$&$15.2^{+0.3}_{-0.3}$&$...$&$...$&$...$&$43.83$\\
Mrk $1044$&B&2.27&$18^{+24}_{-7}$&$0.180^{+0.001}_{-0.001}$&$21.8^{+0.2}_{-0.2}$&$...$&$...$&$...$&$43.51$\\
IRAS $04416+1215$&A&$2.14^{+0.09}_{-0.10}$&...&...&...&...&...&...&$43.89$\\
SDSS J$074352.02+271239.5$&A&$2.69^{+0.09}_{-0.09}$&...&...&...&...&...&...&$44.85$\\
Mrk $382$&A&$2.66^{+0.01}_{-0.01}$&...&...&...&...&...&...&$43.43$\\
SDSS J$080101.41+184840.7$&B&2.33&$147^{+16}_{-13}$&$0.09^{+0.01}_{-0.01}$&$47^{+26}_{-22}$&$...$&$...$&$...$&$44.16$\\
SDSS J$081441.91+212918.5$&B&2.17&$87^{+9}_{-13}$&$0.18^{+0.08}_{-0.03}$&$17^{+8}_{-7}$&$...$&$...$&$...$&$44.25$\\
PG $0844+349$&A&$2.65^{+0.01}_{-0.01}$&...&...&...&...&...&...&$44.07$\\
SDSS J$085946.35+274534.8$&A&$2.31^{+0.11}_{-0.11}$&...&...&...&...&...&...&$43.92$\\
Mrk $110$&A&$2.25^{+0.01}_{-0.01}$&...&...&...&...&...&...&$44.01$\\
SDSS J$093922.89+370943.9$&B&2.42&$77^{+30}_{-77}$&$0.20^{+0.08}_{-0.03}$&$21\pm 9$&$...$&$...$&$...$&$44.05$\\
PG $0953+414$&B&2.02&$82\pm 6$&$0.35^{+0.38}_{-0.07}$&$9.7^{+2.4}_{-4.2}$&$...$&$...$&$...$&$45.03$\\
SDSS J$100402.61+285535.3$&B&2.31&$117\pm 3$&$0.05^{+0.02}_{-0.01}$&$83^{+285}_{-59}$&$...$&$...$&$...$&$44.89$\\
Mrk $142$&A&$2.54^{+0.04}_{-0.04}$&...&...&...&...&...&...&$43.28$\\
UGC $6728$&A&$1.65^{+0.03}_{-0.03}$&...&...&...&...&...&...&$41.96$\\
PG $1211+143$&B&2.07&$80\pm 1$&$0.6^{+7.1}_{-0.2}$&$5.5^{+1.7}_{-4.9}$&$...$&$...$&$...$&$44.25$\\
IRASF $12397+3333$&C&$2.15$&$229^{+6}_{-5}$&$0.080^{+0.001}_{-0.001}$&$121^{+13}_{-10}$&$0.62^{+0.09}_{-0.10}$&$0.67^{+0.05}_{-0.05}$&$0.67^{+0.02}_{-0.02}$ &$43.59$\\
NGC $4748$&A&$2.61^{+0.01}_{-0.01}$&...&...&...&...&...&...&$42.95$\\
Mrk $493$&B&2.18&$57^{+5}_{-8}$&$0.24\pm 0.02$&$14\pm 1$&$...$&$...$&$...$&$43.42$ \\
KA $1858+4850$&A&$2.32^{+0.12}_{-0.12}$&...&...&...&...&...&...&$43.55$\\
PG $2130+099$&A&$2.64^{+0.01}_{-0.01}$&...&...&...&...&...&...&$43.97$\\
\midrule
\multicolumn{10}{c}{\hbox{Sub-Eddington} Subsample ($\rm\dot{\mathscr{M}}< 3$)} \\
 \midrule
PG $0026+129$&B&1.75&$76^{+11}_{-13}$&$0.35^{+0.09}_{-0.05}$&$13\pm 2$&$...$&$...$&$...$&$44.48$\\
PG $0052+251$&B&1.75&$228^{+6}_{-7}$&$0.070^{+0.002}_{-0.002}$&$151^{+10}_{-11}$&$...$&$...$&$...$&$44.79$\\
Fairall $9$&A&$2.35^{+0.01}_{-0.01}$&...&...&...&...&...&...&$44.26$\\
Ark $120$&B&1.96&$181\pm 2$&$0.080^{+0.001}_{-0.001}$&$78\pm 2$&$...$&$...$&$...$&$44.19$\\
MCG $+08-11-011$&A&$1.71^{+0.01}_{-0.01}$&...&...&...&...&...&...&$43.50$\\
Mrk $374$&A&$2.18^{+0.04}_{-0.04}$&...&...&...&...&...&...&$43.30$\\
Mrk $79$&B&1.86&$353\pm 6$&$0.080^{+0.001}_{-0.001}$&$507^{+19}_{-17}$&$...$&$...$&$...$&$43.54$\\
PG $0804+761$&B&2.08&$60^{+4}_{-5}$&$0.28^{+0.03}_{-0.02}$&$12\pm 1$&$...$&$...$&$...$&$44.34$\\
NGC $2617$&A&$2.13^{+0.01}_{-0.01}$&...&...&...&...&...&...&$43.39$\\
SBS $1116+583A$&B&1.72&$74^{+4}_{-3}$&$0.99^{+0.99}_{-0.49}$&$4.7^{+2.9}_{-0.1}$&$...$&$...$&$...$&$42.65$\\
Arp $151$&A&$1.76^{+0.04}_{-0.04}$&...&...&...&...&...&...&$42.82$\\
MCG $+06-26-012$&A&$2.19^{+0.07}_{-0.07}$&...&...&...&...&...&...&$42.92$\\
Mrk $1310$&A&$2.08^{+0.05}_{-0.05}$&...&...&...&...&...&...&$42.88$\\
Mrk $766$&C&1.94& $83\pm 1$&$0.6^{+0.5}_{-0.1}$&$7.5^{+1.7}_{-3.1}$&$0.98^{+0.08}_{-0.13}$&$0.77^{+0.09}_{-0.06}$&$0.62^{+0.02}_{-0.03}$ &$42.96$\\
Ark $374$&A&$2.53^{+0.01}_{-0.01}$&...&...&...&...&...&...&$43.75$\\
NGC $4593$&B&1.67&$141^{+7}_{-10}$&$0.080^{+0.002}_{-0.001}$&$222^{+26}_{-19}$&$...$&$...$&$...$&$42.53$\\
PG $1307+085$&A&$2.28^{+0.03}_{-0.03}$&...&...&...&...&...&...&$43.94$\\
Mrk $279$&B&1.82&$15^{+10}_{-14}$&$0.29^{+0.02}_{-0.01}$&$12.8^{+0.5}_{-0.5}$&$...$&$...$&$...$&$43.85$\\
NGC $5548$&C&1.63& $388\pm 2$&$0.080^{+0.001}_{-0.001}$&$181^{+10}_{-7}$&$6.5^{+0.6}_{-0.5}$&$2.07^{+0.01}_{-0.02}$&$0.47^{+0.01}_{-0.01}$ &$43.33$\\
PG $1426+015$&A&$2.51^{+0.01}_{-0.01}$&...&...&...&...&...&...&$44.46$\\
Mrk $817$&A&$2.46^{+0.01}_{-0.01}$&...&...&...&...&...&...&$43.71$\\
Mrk $290$&C&1.52& $92\pm 3$&$0.6^{+6.0}_{-0.3}$&$3.6^{+2.5}_{-3.2}$&$2.3^{+0.4}_{-0.3}$&$1.3\pm 0.1$&$0.70^{+0.02}_{-0.03}$ &$43.28$\\
Mrk $876$&A&$2.32^{+0.02}_{-0.02}$&...&...&...&...&...&...&$44.35$\\
NGC $6814$&A&$1.72^{+0.01}_{-0.01}$&...&...&...&...&...&...&$42.00$\\
Mrk $509$&B&1.80&$72\pm 1$&$0.9^{+0.9}_{-0.2}$&$5.0^{+1.1}_{-1.9}$&$...$&$...$&$...$&$44.21$\\
NGC $7469$&C&1.89& $88\pm 1$&$0.41\pm 0.03$&$9.0^{+0.5}_{-0.6}$&$0.14^{+0.01}_{-0.02}$&$2.36^{+0.04}_{-0.04}$&$1$ (fixed) &$43.40$ \\
\enddata
\tablecomments{Column (1): object name. 
Column (2): {\sc xspec} spectral fitting model. A: {\sc phabs*zpowerlw}; B: {\sc phabs*(comptt+zpowerlw)}; C: 
{\sc phabs*zxipcf*(comptt+zpowerlw)}. In Model B or C, the {\sc zpowerlw} component is fixed to that constrained from the \hbox{rest-frame} $>2$ keV spectral fitting (see Table~\ref{tbl2}).
Column (3): power-law photon index. It is tied to the $\Gamma$ value in Table~\ref{tbl2} for Model B or C.
Columns (4)--(6): parameters of the {\sc comptt} component (Model B or C), including the temperature of seed photons, 
temperature of electrons in the warm corona, and optical depth of the warm corona.   
Columns (7)--(9): column density, ionization state, and covering factor of the partial-covering ionized absorption component
({\sc zxipcf}; Model C). 
Column (10): logarithm of the rest-frame \hbox{0.3--2 keV} X-ray luminosity.}
\label{tbl3}
\end{deluxetable*}
\end{longrotatetable}

\end{document}